\documentclass[12pt]{article}
\usepackage{amssymb, amsmath, graphicx}

\def\R{\mathbb{R}}
\def\C{\mathbb{C}}
\def\N{\mathbb{N}}

\def\Z{\mathbb{Z}}
\def\i{{\rm 1\kern -.3600em 1}}

\newtheorem{theorem}{Theorem}%[section]
\newtheorem{corollary}[theorem]{Corollary}
\newtheorem{lemma}[theorem]{Lemma}
\newtheorem{proposition}[theorem]{Proposition}

\newtheorem{example}[theorem]{Example}
\newtheorem{remark}[theorem]{Remark}

\begin{document}

\author{\textbf{Dmitri L.~Finkelshtein} \\
{\small Institute of Mathematics, National Academy of Sciences of Ukraine,
Kiev, Ukraine}\\
{\small fdl@imath.kiev.ua} \and
\textbf{Yuri G.~Kondratiev} \\
{\small Fakult\"at f\"ur Mathematik, Universit\"at Bielefeld,
D 33615 Bielefeld, Germany}\\
{\small Forschungszentrum BiBoS, Universit\"at Bielefeld,
D 33615 Bielefeld, Germany}\\
{\small National University ``Kyiv-Mohyla Academy'', Kiev, Ukraine}\\
{\small kondrat@mathematik.uni-bielefeld.de} \and
\textbf{Maria Jo\~{a}o Oliveira} \\
{\small Universidade Aberta, P 1269-001 Lisbon, Portugal}\\
{\small Centro de Matem\'atica e Aplica{\c c}\~oes Fundamentais,}\\
{\small University of Lisbon, P 1649-003 Lisbon, Portugal}\\
{\small Forschungszentrum BiBoS, Universit\"at Bielefeld,
D 33615 Bielefeld, Germany}\\
{\small oliveira@cii.fc.ul.pt}}

\title{Markov evolutions and hierarchical equations in the continuum I.~One-component systems}

\date{}
\maketitle

\vspace*{-1cm}

\begin{abstract}
General birth-and-death as well as hopping stochastic dynamics of infinite 
particle systems in the continuum are considered. We derive corresponding 
evolution equations for correlation functions and generating functionals. 
General considerations are illustrated in a number of concrete examples of 
Markov evolutions appearing in applications.
\end{abstract}


%\medskip

%\noindent
%\textbf{MSC Classification:}

\newpage

\tableofcontents

\newpage
\section{Introduction}

The theory of stochastic lattice gases on the cubic lattice $\Z^d$,
$d\in\N$, is one of the most well developed areas in the interacting
particle systems theory. In the lattice gas models with spin space
$S=\{0,1\}$, the configuration space is defined as $\mathcal{X}=
\{0,1\}^{\Z^d}$. Given a configuration $\sigma=\{\sigma(x):
x\in\Z^d\}\in \mathcal{X}$, we say that a lattice site $x\in\Z^d$ is
free or occupied by a particle depending on $\sigma(x)=0$ or
$\sigma(x)=1$, respectively. The spin-flip dynamics of such a system
means that, at each site $x$ of the lattice, a particle randomly
appears (if the site $x$ is free) or disappears from that site. The
generator of this dynamics is given by
\[
(Lf)(\sigma)=\sum_{x\in\Z^d}a(x,\sigma)(f(\sigma^x)-f(\sigma)),
\]
where $\sigma^x$ denotes the configuration $\sigma$ in which a
particle located at $x$ has disappeared or a new particle has
appeared at $x$. Hence, this dynamics may be interpreted as a
birth-and-death process on $\Z^d$. An example of such a type of
process is given by the classical contact model, which describes the
spread of an infectious disease. In this model an individual at
$x\in\Z^d$ is infected if $\sigma(x)=1$ and healthy if
$\sigma(x)=0$. Healthy individuals become infected at a rate which
is proportional to the number of infected neighbors
($\lambda\sum_{y:|y-x|=1}\sigma(y)$, for some $\lambda\geq 0$),
while infected individuals recover at a rate identically equal to 1.
An additional example is the linear voter model, in which an
individual located at a $x\in\Z^d$ has one of two possible positions
on an issue. He reassesses his view by the influence of surrounding
people. Further examples of such a type may be found e.g.
in~\cite{Lig85}, \cite{Lig99}.

In all these examples clearly there is no conservation on the number
of particles involved. In contrast to them, in the spin-exchange
dynamics there is conservation on the number of particles. In this
case, particles randomly hop from one site in $\Z^d$ to another
one. The generator of such a dynamics is given by
\[
(Lf)(\sigma)= \sum_{x\in\Z^d}\sum_{y\in\Z^d:|y-x|=1}c(x,y,\sigma)(f(\sigma^{xy})-f(\sigma)),
\]
where $\sigma^{xy}$ denotes the configuration $\sigma$ in which a particle
located at $x$ hops to a site $y$.

In this work we consider continuous particle systems, i.e., systems
of particles which can be located at any site in the Euclidean space
$\R^d$, $d\in\N$. In this case, the configuration space of such
systems is the space $\Gamma$ of all locally finite subsets of
$\R^d$. Thus, an analog of the above mentioned spin-flip dynamics
should be a process in which particles randomly appear or disappear
from the space $\R^d$, i.e., a spatial birth-and-death process. The
generator of such a process is informally given by
\begin{eqnarray*}
(LF)(\gamma)&=&\sum_{x\in\gamma}d(x,\gamma\setminus\{x\})
\left(F(\gamma\setminus\{x\})-F(\gamma)\right)\\
&&+ \int_{\R^d}dx\, b(x,\gamma)\left(F(\gamma\cup\{x\})-F(\gamma)\right),
\end{eqnarray*}
where the coefficient $d(x,\gamma)$ indicates the rate at which a particle
located at $x$ in a configuration $\gamma$ dies or disappears, while
$b(x,\gamma)$ indicates the rate at which, given a configuration $\gamma$, a
new particle is born or appears at a site $x$.

By analogy, one may also consider a continuous version of the contact and
voter models above presented. Both continuous versions yield a similar
informal expression for the corresponding generators.

Moreover, one may also consider the analog of the spin-exchange
dynamics. We consider a general case of hopping particle systems, in
which particles randomly hop over the space $\R^d$. In terms of
generators, this means that the dynamics is informally given by
\[
(LF)(\gamma)= \sum_{x\in\gamma}\int_{\R^d}dy\,c(x,y,\gamma)
\left(F(\gamma\setminus\{x\}\cup\{y\})-F(\gamma)\right),
\]
where the coefficient $c(x,y,\gamma)$ indicates the rate at which a particle
located at $x$ in a configuration $\gamma$ hops to a site $y$.

Spatial birth-and-death processes were first discussed by C.~Preston in
\cite{Pr75}. Under some conditions on the birth and death rates, $b$ and $d$,
the author has proved the existence of such processes in a bounded volume on
$\R^d$. In this case, although the number of particles can be arbitrarily
large,
at each moment of time the total number of particles is always finite. Later
on, the problem of convergence of these processes to an equilibrium one was
analyzed in \cite{L81}, \cite{M89}.

Problems of existence, construction, and uniqueness of spatial
birth-and-death processes in an infinite volume were initiated by R.~A.~Holley
and D.~W.~Stroock in \cite{HS78} for a special case of neighbor
birth-and-death processes on the real line. An extension of the uniqueness
result stated therein may be found in \cite{CR79}.

E.~Gl{\"o}tzl analyzed in \cite{G81}, \cite{G82} the birth-and-death
and the hopping dynamics of continuous particle systems for which a
Gibbs measure $\mu$ is reversible. Although he could not prove the
existence of such processes, he has identified the conditions on the
coefficients, $b,d$ and $c$ under which the corresponding generators
are symmetric operators on the space $L^2(\mu)$. For the particular
case of the Glauber stochastic dynamics, such a process was
effectively constructed in \cite{LK03}. The procedure used therein
was extended in \cite{KLRII05} to a general case of birth-and-death
dynamics and to the hopping dynamics. Recently, in \cite{KS06} the
authors have proved the existence of a contact process. Further
details concerning all these constructions are properly archived
throughout this present work.

In this work we propose an alternative approach for the study of a
dynamics based on combinatorial harmonic analysis techniques on
configuration spaces. This particular standpoint of configuration
space analysis was introduced and developed in \cite{KoKu99},
\cite{K00} (Subsection \ref{Subsection21}). For this purpose, we
assume that the coefficients $b, d$ and $c$ are of the type
\begin{equation}
a(x,\gamma)=\sum_{{\eta \subset \gamma}\atop{\vert\eta\vert < \infty}}A_x(\eta),\ a=b,d,\quad c(x,y,\gamma)=\sum_{{\eta \subset \gamma}\atop{\vert\eta\vert < \infty}}C_{x,y}(\eta),\label{Dt4}
\end{equation}
respectively. This special form of the coefficients allows the
used of harmonic analysis techniques, namely, the specific ones
yielding from the natural relations between states, observables,
correlation measures, and correlation functions (Subsection
\ref{Subsection22}). Usually, the starting point for the
construction of a dynamics is the Markov generator $L$ related to
the Kolmogorov equation
\[
\frac{\partial}{\partial t}F_t=LF_t.
\]
Given an initial distribution $\mu$ of the system (from a set of
admissible initial distributions on $\Gamma$), the generator $L$
determines a Markov process on $\Gamma$ which initial distribution
is $\mu$. In alternative to this approach, the natural relations
between observables (i.e., functions defined on $\Gamma$), states,
correlation measures, and correlation functions yield a description
of the underlying dynamics in terms of those elements (Subsection
\ref{Subsection22}), through corresponding Kolmogorov equations.
Such equations are presented under quite general assumptions,
sufficient to define these equations. However, let us observe that on
each concrete application the explicit form of the rates determines
specific assumptions, which only hold for that concrete application.
Such an analysis is discussed separately. In Subsection
\ref{Subsection23} we widen the dynamical description towards the
Bogoliubov functionals \cite{Bog46}, cf.~\cite{KoKuOl02}.

Let us underlying that assumptions (\ref{Dt4}) are natural and quite
general. As a matter of fact, the birth and death rates on the
Glauber, the contact model and the linear and polynomial
voter models dynamics, are both of this type (Subsections
\ref{Subsubsection321}--\ref{Subsubsection324}), as well the
coefficient $c$ for the Kawasaki dynamics (Subsection
\ref{Subsubsection421}).

From the technical point of view, the procedure that is presented here turns
out to be an effective method for the study of equilibrium and non-equilibrium
problems for infinite particle systems in the continuum. This has been
recently emphasized in the construction of a non-equilibrium Glauber dynamics
done in \cite{KoKtZh06}, cf.~considerations at the end of Subsection
\ref{Subsubsection321}.

In our forthcoming publication \cite{FKO06} we present an extension of this
technique towards multicomponent systems. In particular, it yields a new
approach to the study of, e.g., conflict, predator-prey, and Potts-Kawasaki
models.

\section{Markov evolutions in configuration spaces\label{Section2}}

\subsection{Harmonic analysis on configuration spaces\label{Subsection21}}

The configuration space $\Gamma :=\Gamma _{\mathbb{R}^d}$ over $\mathbb{R}^d$,
$d\in\mathbb{N}$, is defined as the set of all locally finite subsets of
$\mathbb{R}^d$,
\[
\Gamma :=\left\{ \gamma \subset \mathbb{R}^d:\left| \gamma_\Lambda\right|
<\infty \hbox{
for every compact }\Lambda\subset \mathbb{R}^d\right\} ,
\]
where $\left| \cdot \right|$ denotes the cardinality of a set and
$\gamma_\Lambda := \gamma \cap \Lambda$. As usual we identify each
$\gamma \in \Gamma $ with the non-negative Radon measure $\sum_{x\in
\gamma }\delta_x\in \mathcal{M}(\mathbb{R}^d)$, where $\delta_x$ is
the Dirac measure with unit mass at $x$,
$\sum_{x\in\emptyset}\delta_x$ is, by definition, the zero measure,
and $\mathcal{M}(\mathbb{R}^d)$ denotes the space of all
non-negative Radon measures on the Borel $\sigma$-algebra
$\mathcal{B}(\mathbb{R}^d)$. This identification allows to endow
$\Gamma $ with the topology induced by the vague topology on
$\mathcal{M}(\mathbb{R}^d)$, i.e., the weakest topology on $\Gamma$
with respect to which all mappings
\[
\Gamma \ni \gamma \longmapsto \langle f,\gamma\rangle :=
\int_{\mathbb{R}^d}d\gamma(x)\,f(x)=\sum_{x\in \gamma }f(x),\quad
f\in C_c(\R^d),
\]
are continuous. Here $C_c(\R^d)$ denotes the set of all continuous functions
on $\R^d$ with compact support. We denote by $\mathcal{B}(\Gamma )$ the
corresponding Borel $\sigma$-algebra on $\Gamma$.

Let us now consider the space of finite configurations
\[
\Gamma_0 := \bigsqcup_{n=0}^\infty \Gamma^{(n)},
\]
where $\Gamma^{(n)} := \Gamma^{(n)}_{\R^d} := \{ \gamma\in \Gamma:
\vert \gamma\vert = n\}$ for $n\in \N$ and $\Gamma^{(0)} :=
\{\emptyset\}$. For $n\in \N$, there is a natural bijection between
the space $\Gamma^{(n)}$ and the symmetrization
$\widetilde{(\R^d)^n}\diagup S_n$ of the set $\widetilde{(\R^d)^n}:=
\{(x_1,...,x_n)\in (\R^d)^n: x_i\not= x_j \hbox{ if } i\not= j\}$
under the permutation group $S_n$ over $\{1,...,n\}$ acting on
$\widetilde{(\R^d)^n}$ by permuting the coordinate indexes. This
bijection induces a metrizable topology on $\Gamma^{(n)}$, and we
endow $\Gamma_0$ with the topology of disjoint union of topological
spaces. By $\mathcal{B}(\Gamma^{(n)})$ and $\mathcal{B}(\Gamma_0)$
we denote the corresponding Borel $\sigma$-algebras on
$\Gamma^{(n)}$ and $\Gamma_0$, respectively.

We proceed to consider the $K$-transform \cite{Le72}, \cite{Le75a},
\cite{Le75b}, \cite{KoKu99}, that is, a mapping which maps functions
defined on $\Gamma_0$ into functions defined on the space $\Gamma$.
Let $\mathcal{B}_c(\R^d)$ denote the set of all bounded Borel sets
in $\R^d$, and for any $\Lambda\in \mathcal{B}_c(\R^d)$ let
$\Gamma_\Lambda := \{\eta\in \Gamma: \eta\subset \Lambda\}$.
Evidently $\Gamma_\Lambda = \bigsqcup_{n=0}^\infty
\Gamma_\Lambda^{(n)}$, where $\Gamma_\Lambda^{(n)}:= \Gamma_\Lambda
\cap \Gamma^{(n)}$ for each $n\in \N_0$, leading to a situation
similar to the one for $\Gamma_0$, described above. We endow
$\Gamma_\Lambda$ with the topology of the disjoint union of
topological spaces and with the corresponding Borel $\sigma$-algebra
$\mathcal{B}(\Gamma_\Lambda)$.

Given a $\mathcal{B}(\Gamma_0)$-measurable function $G$ with local
support, that is, $G\!\!\upharpoonright
_{\Gamma\backslash\Gamma_\Lambda}\equiv 0$ for some $\Lambda \in
\mathcal{B}_c(\mathbb{R}^d)$, the $K$-transform of $G$ is a mapping
$KG:\Gamma\to\mathbb{R}$ defined at each $\gamma\in\Gamma$ by
\begin{equation}
(KG)(\gamma ):=\sum_{{\eta \subset \gamma}\atop{\vert\eta\vert < \infty} }
G(\eta ).
\label{Eq2.9}
\end{equation}
Note that for every such function $G$ the sum in (\ref{Eq2.9}) has only a
finite number of summands different from zero, and thus $KG$ is a well-defined
function on $\Gamma$. Moreover, if $G$ has support described as before, then
the restriction $(KG)\!\!\upharpoonright _{\Gamma _\Lambda }$ is a
$\mathcal{B}(\Gamma_\Lambda)$-measurable function and
$(KG)(\gamma)=(KG)\!\!\upharpoonright _{\Gamma _\Lambda }\!\!(\gamma_\Lambda)$
for all $\gamma\in\Gamma$, i.e., $KG$ is a cylinder function.

Let now $G$ be a bounded $\mathcal{B}(\Gamma_0)$-measurable function
with bounded support, that is, $G\!\!\upharpoonright _{\Gamma
_0\backslash \left(\bigsqcup_{n=0}^N\Gamma _\Lambda ^{(n)}\right)
}\equiv 0$ for some $N\in\N_0, \Lambda \in \mathcal{B}_c(\R^d)$. In
this situation, for each $C\geq \vert G\vert$ one finds $\vert
(KG)(\gamma)\vert\leq C(1+\vert\gamma_\Lambda\vert)^N$ for all
$\gamma\in\Gamma$. As a result, besides the cylindricity property,
$KG$ is also polynomially bounded. In the sequel we denote the space
of all bounded $\mathcal{B}(\Gamma_0)$-measurable functions with
bounded support by $B_{bs}(\Gamma_0)$. It has been shown in
\cite{KoKu99} that the $K$-transform is a linear isomorphism which
inverse mapping is defined on cylinder functions by
\[
\left( K^{-1}F\right) (\eta ):=\sum_{\xi \subset \eta }(-1)^{|\eta
\backslash \xi |}F(\xi ),\quad \eta \in \Gamma _0.
\]
As a side remark, we observe that this property of the $K$-transform
yields a full complete description of the elements in
$\mathcal{FP}(\Gamma):=K\left(B_{bs}(\Gamma_0)\right)$ which may be
found in \cite{KoKu99}, \cite{KoKuOl00b}. However, throughout this
work we shall only make use of the above described cylindricity and
polynomial boundedness properties of the functions in
$\mathcal{FP}(\Gamma)$.

Among the elements in the domain of the $K$-transform are also the so-called
coherent states $e_\lambda(f)$ corresponding to
$\mathcal{B}(\mathbb{R}^d)$-measurable functions $f$ with compact support. By
definition, for any $\mathcal{B}(\mathbb{R}^d)$-measurable function $f$,
\[
e_\lambda (f,\eta ):=\prod_{x\in \eta }f(x) ,\ \eta \in
\Gamma _0\!\setminus\!\{\emptyset\},\quad  e_\lambda (f,\emptyset ):=1.
\]
If $f$ has compact support, then the image of $e_\lambda (f)$
under the $K$-transform is a function on $\Gamma$ given by
\[
\left( Ke_\lambda (f)\right) (\gamma )=\prod_{x\in \gamma }(1+f(x)),\quad
\gamma\in \Gamma.
\]

As well as the $K$-transform, its dual operator $K^*$ will also play
an essential role in our setting. Let
$\mathcal{M}_{\mathrm{fm}}^1(\Gamma)$ denote the set of all
probability measures $\mu$ on $(\Gamma ,\mathcal{B}(\Gamma))$ with
finite local moments of all orders, i.e.,
\begin{equation}
\int_\Gamma d\mu (\gamma)\, |\gamma _\Lambda |^n<\infty\quad
\mathrm{for\,\,all}\,\,n\in\N
\mathrm{\,\,and\,\,all\,\,} \Lambda \in \mathcal{B}_c(\R^d).\label{Dima5}
\end{equation}
By the definition of a dual operator, given a
$\mu\in\mathcal{M}_{\mathrm{fm}}^1(\Gamma)$, the so-called correlation
measure $\rho_\mu:=K^*\mu$ corresponding to $\mu$ is a measure on
$(\Gamma _0,\mathcal{B}(\Gamma _0))$ defined for each
$G\in B_{bs}(\Gamma_0)$ by
\begin{equation}
\int_{\Gamma _0}d\rho _\mu(\eta )\,G(\eta )=\int_\Gamma d\mu (\gamma)\, \left(
KG\right) (\gamma).  \label{Eq2.16}
\end{equation}
Observe that under the above conditions $K\!\left|G\right|$ is
$\mu$-integrable. In terms of correlation measures this
means that $B_{bs}(\Gamma_0)\subset L^1(\Gamma_0,\rho_\mu)$.

Actually, $B_{bs}(\Gamma_0)$ is dense in $L^1(\Gamma_0,\rho_\mu)$.
Moreover, still by (\ref{Eq2.16}), on $B_{bs}(\Gamma_0)$ the inequality
$\Vert KG\Vert_{L^1(\mu)}\leq \Vert G\Vert_{L^1(\rho_\mu)}$
holds, allowing then an extension of the $K$-transform to a bounded operator
$K:L^1(\Gamma_0,\rho_\mu)\to L^1(\Gamma,\mu)$ in such a way that equality
(\ref{Eq2.16}) still holds for any $G\in L^1(\Gamma_0,\rho_\mu)$. For the
extended operator the explicit form (\ref{Eq2.9}) still holds, now $\mu$-a.e.
This means, in particular,
\begin{equation}
\left( Ke_\lambda (f)\right) (\gamma) = \prod_{x\in \gamma }(1+f(x)),\quad
\mu \mathrm{-a.a.}\,\gamma\in\Gamma,\label{1.1}
\end{equation}
for all $\mathcal{B}(\R^d)$-measurable functions $f$ such that
$e_\lambda(f)\in L^1(\Gamma_0,\rho_\mu)$, cf.~e.g. \cite{KoKu99}.

We also note that in terms of correlation measures $\rho_\mu$ property
(\ref{Dima5}) means that $\rho_\mu$ is locally finite, that is,
$\rho_\mu(\Gamma _\Lambda ^{(n)})<\infty$ for all $n\in\N_0$ and all
$\Lambda\in\mathcal{B}_c(\R^d)$. By $\mathcal{M}_{\mathrm{lf}}(\Gamma_0)$ we
denote the class of all locally finite measures on $\Gamma_0$.

\begin{example}
\label{example}
Given a constant $z>0$, let $\pi_z$ be the Poisson measure with
intensity $zdx$, that is, the probability measure on
$(\Gamma,\mathcal{B}(\Gamma))$ with Laplace transform given by
\[
\int_\Gamma d\pi_z(\gamma )\,\exp \left( \sum_{x\in \gamma }\varphi (x)\right)
=\exp \left( z\int_{\R^d}dx\,\left( e^{\varphi (x)}-1\right)\right)
\]
for all $\varphi\in\mathcal{D}$. Here $\mathcal{D}$ denotes the Schwartz space
of all infinitely differentiable real-valued functions on $\R^d$ with compact
support. The correlation measure corresponding to $\pi_z$ is the so-called
Lebesgue-Poisson measure
$$
\lambda_z:=\sum_{n=0}^\infty \frac{z^n}{n!} m^{(n)},
$$
where each $m^{(n)}$, $n\in \N$, is the image measure on $\Gamma^{(n)}$ of
the product measure $dx_1...dx_n$ under the mapping
$\widetilde{(\R^d)^n}\ni (x_1,...,x_n)\mapsto\{x_1,...,x_n\}\in \Gamma^{(n)}$.
For $n=0$ we set $m^{(0)}(\{\emptyset\}):=1$. This special case emphasizes
the technical role of the coherent states in our setting. First,
$e_\lambda(f)\in L^p(\Gamma_0, \lambda_z)$ whenever $f\in L^p(\R^d, dx)$ for
some $p\geq 1$, and, moreover, $\Vert e_\lambda(f)\Vert^p_{L^p(\lambda_z)}
=\exp(z\Vert f\Vert^p_{L^p(dx)})$. Second, given a dense subspace
$\mathcal{L}\subset L^2(\R^d,dx)$, the set $\{e_\lambda(f):f\in \mathcal{L}\}$
is total in $L^2(\Gamma_0,\lambda_z)$.
\end{example}

Given a probability measure $\mu$ on $\Gamma$, let $\mu\circ p_\Lambda ^{-1}$
be the image measure on the space $\Gamma _\Lambda $,
$\Lambda\in\mathcal{B}_c(\R^d)$, under the mapping
$p_\Lambda :\Gamma \rightarrow \Gamma _\Lambda $ defined by
$p_\Lambda (\gamma ):=\gamma _\Lambda $, $\gamma \in \Gamma $, i.e., the
projection of $\mu$ onto $\Gamma _\Lambda$. A measure
$\mu\in\mathcal{M}_{\mathrm{fm}}^1(\Gamma)$ is called locally absolutely
continuous with respect to $\pi:=\pi_1$ whenever for each
$\Lambda\in\mathcal{B}_c(\R^d)$ the measure $\mu\circ p_\Lambda ^{-1}$ is
absolutely continuous with respect to $\pi\circ p_\Lambda ^{-1}$. In this
case, the correlation measure $\rho_\mu$ is absolutely continuous with respect
to the Lebesgue-Poisson measure $\lambda:=\lambda_1$. The Radon-Nikodym
derivative $k_\mu:=\frac{d\rho_\mu}{d\lambda}$ is the so-called correlation
function corresponding to $\mu$. For more details see e.g.~\cite{KoKu99}.

\subsection{Markov generators and related evolutional equations\label{Subsection22}}

Before proceeding further, let us first summarize graphically all the above
described notions as well as their relations (see the diagram below). Having
in mind concrete applications, let us also mention the natural meaning of this
diagram in the context of a given infinite particle system.

%\documentstyle[12pt]{article}

%\begin{document}

\vspace{1truecm}

\begin{center}

\def\xpos {0}
\def\ypos {0}

\def\xlen {200}
\def\ylen {100}

\def\xinit {10}
\def\yinit {10}

% Tamanho das linhas

\def\xlenline {180}           % = \xlen - 2 * \xinit
\def\ylenline {80}           % = \ylen - 2 * \yinit

% Posicao dos vectores inversos

\def\xposvectori {190}        % = \xlen - \xinit
\def\yposvectori {90}        % = \ylen - \yinit

% Posicao das letras

\def\xletraposinit {-9}       % = \xpos - 9
\def\xletraposend {191}       % = \xlen - 9

\def\yletraposinit {-6}       % = \ypos - 6
\def\yletraposend {91}       % = \ylen - 9

% Posicao das expressoes nos vectores

\def\yexprtop {120}           % = \ylen + 20
\def\yexprbottom {-18}        % = \ypos - 18
\def\xexprleft {-14}          % = \xpos - 14
\def\xexprright {203}         % = \xlen + 3

\begin{picture} (\xlen, \ylen) (0, 0)

% Setas

\put (\xinit,       \ypos) {\vector (1,0)  {\xlenline}}
\put (\xposvectori, \ypos) {\vector (-1,0) {\xlenline}}

\put (\xpos, \yinit)       {\vector (0,1) {\ylenline}}

\put (\xinit, \ylen)       {\vector (1,0) {\xlenline}}
\put (\xposvectori, \ylen) {\vector (-1,0) {\xlenline}}

\put (\xlen, \yposvectori) {\vector (0,-1) {\ylenline}}

% Letras dos cantos

\put (\xletraposinit, \yletraposinit) {\makebox(20,12)[t] {$G$}}
\put (\xletraposinit, \yletraposend)  {\makebox(20,12)[t]{$F$}}
\put (\xletraposend,  \yletraposend)  {\makebox(20,12)[t]{$\mu$}}
\put (\xletraposend,  \yletraposinit) {\makebox(20,12)[t]{$\rho _{_{}\mu}$}}

% Expressoes

\put (\xpos, \yexprtop) 
{\makebox(\xlen, 12)[t] 
{$\langle F,\mu\rangle = \displaystyle\int_{\Gamma}d\mu (\gamma) F(\gamma)$}}

\put (\xpos, \yexprbottom)
{\makebox(\xlen, 12)[t] {$\langle G,\rho _{_{}\mu}\rangle = \displaystyle\int_{\Gamma_0}d\rho _{_{}\mu}(\eta) G(\eta)$}}

\put (\xexprleft, \ypos) 
{\makebox(12, \ylen)[l] {$K$}}

\put (\xexprright, \ypos) 
{\makebox(14, \ylen)[l] {$K^{*}$}}

\end{picture}

\end{center}

\vspace{1truecm}

%\end{document}

The state of such a system is described by a probability measure $\mu$ on
$\Gamma$ and the functions $F$ on $\Gamma$ are considered as observables of
the system. They represent physical quantities which can be measured. The
expected values of the measured observables correspond to the expectation
values $\langle F,\mu\rangle:=\int_\Gamma d\mu(\gamma)\,F(\gamma)$.

In this interpretation we call the functions $G$ on $\Gamma_0$
quasi-observables, because they are not observables themselves, but they can
be used to construct observables via the $K$-transform. In this way we obtain
all observables which are additive in the particles, namely, energy, number of
particles.

The description of the underlying dynamics of such a system is
an essentially interesting and often a difficult question. The number of
particles involved, which imposes a natural complexity to the study, on the
one hand, and the infinite dimensional analysis methods and tools available,
once in a while either limited or insufficient, on the other hand, are
physical and mathematical reasons for the difficulties, and failures, pointed
out. However, it arises from the previous diagram an alternative approach
to the construction of the dynamics, overcoming some of those difficulties.

As usual the starting point for this approach is the Markov generator of the
dynamics, in the sequel denoted by $L$, related to the Kolmogorov equation for
observables
\[
\frac{\partial}{\partial t}F_t=LF_t.\eqno (\mathrm{KE})
\]
Given an initial distribution $\mu$ of the system (from a set of admissible
initial distributions on $\Gamma$), the generator $L$ determines a Markov
process on $\Gamma$ which initial distribution is $\mu$. Within the diagram
context, the distribution $\mu_t$ of the Markov process at each time $t$ is
then a solution of the dual Kolmogorov equation
\[
\frac d{dt}\mu_t=L^*\mu_t,\eqno (\mathrm{KE})^*
\]
$L^*$ being the dual operator of $L$.

The use of the $K$-transform allows us to proceed further. As a matter of
fact, if $L$ is well-defined for instance on $\mathcal{FP}(\Gamma)$, then
its image under the $K$-transform $\hat L:=K^{-1}LK$ yields a Kolmogorov
equation for quasi-observables
\[
\frac {\partial}{\partial t}G_t=\hat LG_t.\eqno (\mathrm{QKE})
\]
Through the dual relation between quasi-observables and correlation measures
this leads naturally to a time evolution description of the correlation
function $k_\mu$ corresponding to the initial distribution $\mu$ given above.
Of course, in order to obtain such a description we must assume that at each
time $t$ the correlation measure corresponding to the distribution $\mu_t$ is
absolutely continuous with respect to the Lebesgue-Poisson measure
$\lambda$. Then, denoting by $\hat L^*$ the dual operator of $\hat L$ in the
sense
\[
\int_{\Gamma_0}d\lambda(\eta)\,(\hat LG)(\eta) k(\eta)=
\int_{\Gamma_0}d\lambda(\eta)\,G(\eta) (\hat L^*k)(\eta),
\]
one derives from (QKE) its dual equation,
\[
\frac {\partial}{\partial t}k_t=\hat L^*k_t.\eqno (\mathrm{QKE})^*
\]
Clearly, the correlation function $k_t$ corresponding to $\mu_t$,
$t\geq 0$, is a solution of $(\mathrm{QKE})^*$. At this point it is
opportune to underline that a solution of $(\mathrm{QKE})^*$ does
not have to be a correlation function (corresponding to some measure
on $\Gamma$), a fact which is frequently not taken into account in
theoretical physics discussions. An additional analysis is needed in
order to distinguish the correlation functions from the set of
solutions of the $(\mathrm{QKE})^*$ equation. Within our setting,
some criteria were developed in \cite{BeKoKuLy99}, \cite{Le75b}, \cite{KoKu99},
\cite{K00}.

In this way we have derived four equations related to the dynamics of an
infinite particle system in the continuum. Starting with (KE), one had derived
$(\mathrm{QKE})^*$, both equations being well-known in physics. Concerning the
latter equation, let us mention its Bogoliubov hierarchical structure, which
in the Hamiltonian dynamics case yields the well-known BBGKY-hierarchy (see
e.g.~\cite{Bog46}). In our case, the hierarchical structure is given by a
countable infinite system of equations
\begin{equation}
\frac {\partial}{\partial t}k_t^{(n)}=(\hat L^*k_t)^{(n)},\quad
k_t^{(n)}:=k_t\!\!\upharpoonright_{\Gamma^{(n)}},\
(\hat L^*k_t)^{(n)}:=(\hat L^*k_t)\!\!\upharpoonright_{\Gamma^{(n)}},\ n\in\N_0.
\label{nova}
\end{equation}
In contrast to (KE), note that each equation in (\ref{nova}) only depends on a
finite number of coordinates. This explains the technical efficacy of equation
$(\mathrm{QKE})^*$ in concrete applications.

Although equations (QKE) and $(\mathrm{KE})^*$ being also
known in physics, their studied is not so developed and usually they are not
exploit in concrete applications. However, in such applications those
equations often turn out to be an effective method.

Before proceeding to concrete applications, let us observe that for some
concrete models it is possible to widen the dynamical description towards
Bogoliubov functionals \cite{Bog46}.

\subsection{Generating functionals\label{Subsection23}}

Given a probability measure $\mu$ on $(\Gamma, \mathcal{B} (\Gamma))$ the
so-called Bogoliubov or generating functional $B_\mu$ corresponding to $\mu$
is the functional defined at each $\mathcal{B}(\R^d)$-measurable
function $\theta$ by
\begin{equation}
B_\mu(\theta) :=\int_\Gamma d\mu(\gamma)\,\prod_{x\in\gamma}(1+\theta (x)),
\label{Dima2}
\end{equation}
provided the right-hand side exists for $\left|\theta\right|$. In the same way
one cannot define the Laplace transform for all measures on $\Gamma$, it is
clear from (\ref{Dima2}) that one cannot define the Bogoliubov functional for
all probability measures on $\Gamma$ as well. Actually, for each
$\theta >-1$ so that the right-hand side of (\ref{Dima2}) exists, one may equivalently rewrite (\ref{Dima2}) as
\[
B_\mu(\theta) :=\int_\Gamma d\mu(\gamma)\,e^{\langle\ln(1+\theta),\gamma)\rangle},
\]
showing that $B_\mu$ is a modified Laplace transform.

If the Bogoliubov functional $B_\mu$ corresponding to a probability measure
$\mu$ exists, then clearly the domain of $B_\mu$ depends on the
underlying measure. Conversely, the domain of a Bogoliubov functional $B_\mu$
reflects special properties over the measure $\mu$ \cite{KoKuOl02}. For
instance, if $\mu$ has finite local exponential moments, i.e.,
\[
\int_\Gamma d\mu (\gamma )\, e^{\alpha|\gamma _\Lambda |}<\infty \quad
\hbox{for all}\,\,\alpha>0\,\,
\hbox{and all}\,\,\Lambda \in \mathcal{B}_c(\R^d),
\]
then $B_\mu$ is well-defined for instance on all bounded functions $\theta$
with compact support. The converse is also true. In fact, for each $\alpha>0$
and each $\Lambda\in\mathcal{B}_c(\R^d)$ the latter integral is equal to
$B_\mu((e^\alpha-1)\i_\Lambda )$. In this situation, to a such measure $\mu$
one may associate the correlation measure $\rho_\mu$, and equalities
(\ref{Eq2.16}) and (\ref{1.1}) then yield a description of the functional
$B_\mu$ in terms of either the measure $\rho_\mu$:
\[
B_\mu(\theta)
= \int_\Gamma d\mu(\gamma)\,\left( Ke_\lambda (\theta)\right) (\gamma)
= \int_{\Gamma_0}d\rho_\mu(\eta)\, e_\lambda (\theta, \eta),
\]
or the correlation function $k_\mu$, if $\rho_\mu$ is absolutely continuous
with respect to the Lebesgue-Poisson measure $\lambda$:
\[
B_\mu(\theta)=\int_{\Gamma_0}d\lambda(\eta)\, e_\lambda (\theta, \eta)k_\mu(\eta).
\]

Within Subsection \ref{Subsection22} framework, this gives us a way to express
the dynamics of an infinite particle system in terms of the Bogoliubov
functionals
\[
B_t(\theta)=\int_{\Gamma_0}d\lambda(\eta)\,e_\lambda(\theta,\eta)k_t(\eta)
\]
corresponding to the states of the system at each time $t\geq 0$, provided the
functionals exist. Informally,
\begin{equation}
\frac{\partial}{\partial t}B_t(\theta)
=\int_{\Gamma_0} d\lambda(\eta)\,e_\lambda(\theta,\eta)\frac{\partial}{\partial t}k_t(\eta)=\int_{\Gamma_0}d\lambda(\eta)\,(\hat Le_\lambda(\theta))(\eta)k_t(\eta).\label{B1}
\end{equation}
In other words, given the operator $\tilde L$ defined at
\[
B(\theta):=\int_{\Gamma_0}d\lambda(\eta)\,e_\lambda(\theta,\eta)k(\eta)\quad
(k:\Gamma_0\to\R^+_0)
\]
by
\[
(\tilde LB)(\theta):=\int_{\Gamma_0}d\lambda(\eta)\,(\hat Le_\lambda(\theta))(\eta)k(\eta),
\]
heuristically (\ref{B1}) means that the Bogoliubov functionals $B_t$,
$t\geq 0$, are a solution of the equation
\begin{equation}
\frac{\partial}{\partial t}B_t=\tilde LB_t.\label{Dima3}
\end{equation}
Besides the problem of the existence of the Bogoliubov functionals $B_t$,
$t\geq 0$, let us also observe that if a solution of equation (\ref{Dima3})
exists, a priori it does not have to be a Bogoliubov functional corresponding
to some measure. The verification requests an additional analysis, see
e.g.~\cite{KoKuOl02}, \cite{K00}.

In applications below, in order to derive explicit formulas for $\tilde L$,
the next results show to be useful. Here and below, all $L^p_\C$-spaces,
$p\geq 1$, consist of $p$-integrable complex-valued functions.

\begin{proposition}
\label{9Prop9.1.1}Given a measure
$\mu\in\mathcal{M}_{\mathrm{fm}}^1(\Gamma)$ assume that the corresponding
Bogoliubov functional $B_\mu$ is entire on $L^1_\C(\R^d,dx)$. Then each
differential of $n$-th order of $B_\mu$, $n\in\N$, at each
$\theta_0\in L^1_\C(\R^d,dx)$ is defined by a symmetric kernel in
$L^\infty_\C((\R^d)^n,dx_1... dx_n)$ denoted by
$\frac{\delta ^nB_\mu(\theta _0)}{\delta\theta _0(x_1)...\delta\theta _0(x_n)}$
and called the variational derivative of $n$-th order of $B_\mu$ at
$\theta _0$. In other words,
\begin{eqnarray*}
&&\frac{\partial ^n}{\partial z_1...\partial z_n}B_\mu
\left( \theta _0+\sum_{i=1}^nz_i\theta _i\right)\Big\vert_{z_1=...=z_n=0}\\
&=&\int_{\R^d}dx_1\,\theta_1(x_1)\cdots \int_{\R^d}dx_n\,\theta_n(x_n)
\frac{\delta ^nB_\mu(\theta _0)}{\delta \theta _0(x_1)...\delta\theta _0(x_n)},
\end{eqnarray*}
for all $\theta_1,...,\theta_n \in L^1_\C(\R^d,dx)$. Furthermore, using the
notation
\[
\left(D^{\left|\eta\right|} B_\mu\right)(\theta_0;\eta)
:=\frac{\delta ^nB_\mu(\theta _0)}{\delta \theta _0(x_1)...\delta \theta
_0(x_n)}\quad \mathit{for}\mathrm{\,\,}\eta =\{x_1,...,x_n\}\in \Gamma^{(n)}, n\in\N,
\]
the Taylor expansion of $B_\mu$ at each $\theta_0\in L^1_\C(\R^d,dx)$ may be
written in the form
\[
B_\mu(\theta _0+\theta )=\int_{\Gamma _0}d\lambda(\eta)\,
e_\lambda (\theta ,\eta )
\left(D^{\left|\eta\right|} B_\mu\right)(\theta_0;\eta),\quad \theta
\in L^1_{\C}(\R^d,dx).
\]
\end{proposition}

In terms of the measure $\mu$, the holomorphy asssumption in Proposition
\ref{9Prop9.1.1} implies that $\mu$ is locally absolutely continuous with
respect to the measure $\pi$ and the correlation function $k_\mu$ is given for
$\lambda$-a.a~$\eta\in\Gamma_0$ by
$k_\mu(\eta )=\left(D^{\left|\eta\right|} B_\mu\right)(0;\eta)$. Moreover, for
all $\theta\in L^1_\C(\R^d,dx)$ the following relation holds
\begin{equation}
\left(D^{\left|\eta\right|} B_\mu\right)(\theta;\eta)
=\int_{\Gamma _0}d\lambda(\xi)\,k_\mu (\eta \cup \xi )e_\lambda (\theta,\xi),
\quad \lambda-\hbox{a.e.},\label{Dima1}
\end{equation}
showing that the Bogoliubov functional $B_\mu$ is the generating
functional for the correlation functions
$k_\mu\!\!\upharpoonright_{\Gamma^{(n)}}$, $n\in\N_0$. For more details and
proofs see e.g.~\cite{KoKuOl02}.

\subsection{Algebraic properties\label{Subsection24}}

As discussed before, the description of the dynamics of a particle system is
closely related to the operators $L$, $\hat L$, and $\hat L^*$. To explicitly
describe these operators in the examples below, the following algebraic
properties turn out to be powerful tools for a simplification of
calculations.

Given $G_1$ and $G_2$ two $\mathcal{B}(\Gamma_0)$-measurable functions, let us
consider the $\star$-convolution between $G_1$ and $G_2$,
\begin{eqnarray*}
(G_1\star G_2)(\eta )&:=&\sum_{(\eta _1,\eta _2,\eta _3)\in \mathcal{P}_3(\eta)}
G_1(\eta _1\cup\eta _2)G_2(\eta _2\cup\eta _3)\\
&=&\sum_{\xi\subset\eta}G_1(\xi)\sum_{\zeta\subset\xi}G_2((\eta\setminus\xi)\cup\zeta),\quad \eta \in \Gamma _0,
\end{eqnarray*}
where $\mathcal{P}_3(\eta )$ denotes the set of all partitions of $\eta$ in
three parts which may be empty, \cite{KoKu99}. It is straightforward to verify
that the space of all $\mathcal{B}(\Gamma_0)$-measurable functions endowed
with this product has the structure of a commutative algebra with unit element
$e_\lambda(0)$. Furthermore, for every $G_1, G_2\in B_{bs}(\Gamma_0)$ we have
$G_1\star G_2\in B_{bs}(\Gamma_0)$, and
\begin{equation}
K\left( G_1\star G_2\right) =\left( KG_1\right) \cdot \left( KG_2\right)
\label{1.5}
\end{equation}
cf.~\cite{KoKu99}. Concerning the action of the $\star$-convolution on
coherent states one finds
\begin{equation}
e_\lambda(f)\star e_\lambda(g) = e_\lambda(f + g+ fg) \label{2.3}
\end{equation}
for all $\mathcal{B}(\mathbb{R}^d)$-measurable functions $f$ and $g$. More
generally, for all $\mathcal{B}(\Gamma_0)$-measurable functions $G$ and all
$\mathcal{B}(\mathbb{R}^d)$-measurable functions $f$ we have
\begin{equation}
\left(  G\star e_{\lambda }\left( f\right)\right) \left( \eta \right)
=\sum_{\xi \subset \eta }G\left( \xi \right)
e_{\lambda }\left(f+1,\xi \right)e_\lambda\left( f,\eta\setminus \xi \right).\label{Dima}
\end{equation}

Technically the next result shows to be very useful. We refer e.g.~to
\cite{O02} for its proof. In particular, for $n=3$, it yields an integration
result for the $\star$-convolution.

\begin{lemma}
\label{Lmm2}Let $n\in \N$, $n\geq 2$, be given. Then
\begin{eqnarray*}
&&\int_{\Gamma _0}d\lambda(\eta _1)...\int_{\Gamma _0}d\lambda(\eta _n)\,
G(\eta _1\cup ...\cup \eta_n)H(\eta_1,...,\eta _n)
\\
&=&\int_{\Gamma _0}d\lambda(\eta )\,G(\eta )
\sum_{(\eta _1,...,\eta _n)\in \mathcal{P}_n(\eta)}H(\eta _1,...,\eta _n)
\end{eqnarray*}
for all positive measurable functions $G:\Gamma _0\to\R$ and
$H:\Gamma _0\times ...\times \Gamma _0\to\R$. Here
$\mathcal{P}_n(\eta )$ denotes the set of all partitions of $\eta $
in $n$ parts, which may be empty.
\end{lemma}

\begin{lemma}
\label{Lmm3}For all positive measurable functions
$H,G_1,G_2:\Gamma _0\to \R$ one has
\begin{eqnarray*}
&&\int_{\Gamma_0}d\lambda(\eta)\,H(\eta)(G_1\star G_2)(\eta)\\
&=&\int_{\Gamma_0}d\lambda(\eta_1)\int_{\Gamma_0}d\lambda(\eta_2)
\int_{\Gamma_0}d\lambda(\eta_3)H(\eta_1\cup\eta_2\cup\eta_3)
G_1(\eta_1\cup\eta_2)G_2(\eta_2\cup\eta_3).
\end{eqnarray*}
\end{lemma}

\section{Markovian birth-and-death dynamics in configuration spaces\label{Section3}}

In a birth-and-death dynamics, at each random moment of time and at each
site in $\R^d$, a particle randomly appears or disappears according to birth
and death rates which depend on the configuration of the whole system at that
time. Informally, in terms of Markov generators, this behaviour is described
through the operators $D^-_x$ and $D^+_x$ defined at each $F:\Gamma\to\R$ by\footnote{Here and below, for simplicity of notation, we have just written $x$
instead of $\{x\}$.}
\[
(D^-_xF)(\gamma):=F(\gamma\setminus x)-F(\gamma),\quad
(D^+_xF)(\gamma):=F(\gamma\cup x)-F(\gamma),
\]
corresponding, respectively, to the annihilation and creation of a particle at
a site $x$. More precisely,
\begin{equation}
(LF)(\gamma):=\sum_{x\in\gamma}d(x,\gamma\setminus x)
(D^-_xF)(\gamma)
+ \int_{\R^d}dx\, b(x,\gamma)(D^+_xF)(\gamma),\label{M9}
\end{equation}
where the coefficient $d(x,\gamma)\geq 0$ indicates the rate at which a
particle located at $x$ in a configuration $\gamma$ dies or disappears, while
$b(x,\gamma)\geq 0$ indicates the rate at which, given a configuration
$\gamma$, a new particle is born or appears at a site $x$.

\subsection{Markovian birth-and-death generators\label{Subsection31}}

In order to give a meaning to (\ref{M9}) let us consider the class
of measures $\mu\in \mathcal{M}_{\mathrm{fm}}^1(\Gamma)$ such that
$d(x,\cdot),b(x,\cdot)\in L^1(\Gamma,\mu)$, $x\in\R^d$, and for all
$n\in\N_0$ and all $\Lambda\in\mathcal{B}_c(\R^d)$ the following
integrability condition is fulfilled:
\begin{equation}
\int_\Gamma d\mu(\gamma)\,|\gamma _\Lambda |^n\sum_{x\in\gamma_\Lambda}
d(x,\gamma\setminus x)
+\int_\Gamma d\mu(\gamma)\,|\gamma _\Lambda |^n\int_\Lambda dx\,
b(x,\gamma)<\infty.\label{DM5}
\end{equation}
For $F\in\mathcal{FP}(\Gamma)=K( B_{bs}(\Gamma_0))$, this condition
is sufficient to insure that $LF$ is $\mu$-a.e.~well-defined on
$\Gamma$. This follows from the fact that for each $G\in
B_{bs}(\Gamma_0)$ there are $\Lambda\in \mathcal{B}_c(\R^d),
N\in\N_0$ and a $C\geq 0$ such that $G$ has support in $\cup
_{n=0}^N\Gamma _\Lambda ^{(n)}$ and $\vert G\vert\leq C$, which
leads to a cylinder function $F=KG$ such that
$\left|F(\gamma)\right|=\left|F(\gamma_\Lambda)\right|\leq
C(1+\vert\gamma_\Lambda\vert)^N$ for all $\gamma\in\Gamma$
(cf.~Subsection \ref{Subsection21}). Hence (\ref{M9}) and
(\ref{DM5}) imply that $LF\in L^1(\Gamma,\mu)$.

Given a family of functions $B_x, D_x:\Gamma_0\to\R$, $x\in\R^d$,
such that $KB_x\geq 0$, $KD_x\geq 0$, in the following we wish to
consider $KB_x$ and $KD_x$ as birth and death rates, i.e.,
\begin{equation}
b(x,\gamma) = \left(KB_x\right)(\gamma),\ d(x, \gamma)=(KD_x)(\gamma).
\label{Z1}
\end{equation}
We shall then restrict the previous class of measures in
$\mathcal{M}_{\mathrm{fm}}^1(\Gamma)$ to the set of all measures
$\mu\in \mathcal{M}_{\mathrm{fm}}^1(\Gamma)$ such that
$B_x, D_x\in L^1(\Gamma_0,\rho_\mu)$, $x\in\R^d$, and
\begin{equation}
\int_\Gamma d\mu(\gamma)\,|\gamma _\Lambda |^n\left\{\sum_{x\in\gamma_\Lambda}
\left(K\vert D_x\vert\right)(\gamma\setminus x)+\int_\Lambda dx\,
\left(K\vert B_x\vert\right)(\gamma)\right\}<\infty\label{M5}
\end{equation}
for all $n\in\N_0$ and all $\Lambda\in\mathcal{B}_c(\R^d)$. Under
these assumptions, the $K$-transform of each $B_x$ and each $D_x$, $x\in\R^d$,
is well-defined. Moreover, $KB_x,KD_x\in L^1(\Gamma,\mu)$, cf.~Subsection
\ref{Subsection21}. Of course, all previous considerations hold. In addition,
we have the following result for the operator $\hat L$ on quasi-observables.

\begin{proposition}
\label{Prop1} The action of $\hat L$ on functions $G\in B_{bs}(\Gamma_0)$ is
given for $\rho_\mu$-almost all $\eta\in\Gamma_0$ by
\begin{equation}
(\hat LG)(\eta)=
-\sum_{x\in\eta}\left(D_x\star G(\cdot\cup x)\right)(\eta\setminus x)
+\int_{\R^d}dx\,\left(B_x\star G(\cdot\cup x)\right)(\eta).\label{M6}
\end{equation}
Moreover,
$\hat L\left(B_{bs}(\Gamma_0)\right)\subset L^1(\Gamma_0,\rho_\mu)$.
\end{proposition}

\noindent
\textbf{Proof.} By the definition of the $K$-transform, for all
$G\in B_{bs}(\Gamma_0)$ we find
\begin{eqnarray*}
(KG)(\gamma\setminus x)- (KG)(\gamma)&=&-(K(G(\cdot\cup x)))(\gamma\setminus x),\quad x\in\gamma,\\
(KG)(\gamma\cup x)- (KG)(\gamma) &=& (K(G(\cdot\cup x)))(\gamma),\quad x\notin\gamma.
\end{eqnarray*}
Given a $F\in\mathcal{FP}(\Gamma)$ of the form $F=KG$, $G\in
B_{bs}(\Gamma_0)$, these equalities combined with the algebraic
action (\ref{1.5}) of the $K$-transform yield
\begin{eqnarray*}
(LF)(\gamma)&=&-\sum_{x\in\gamma}d(x,\gamma\setminus x)
\left(K\left(G(\cdot\cup x)\right)\right)(\gamma\setminus x)\\
&&+ \int_{\{x:x\notin\gamma\}}\!\!\!dx\,b(x,\gamma)\left(K\left(G(\cdot\cup x)\right)\right)(\gamma)\\
&=&-\sum_{x\in\gamma}\left(K\left(D_x\star G(\cdot\cup x)\right)\right)
(\gamma\setminus x)
+ \int_{\R^d}\!\!\!dx\left(K\left(B_x\star G(\cdot\cup x)\right)\right)
(\gamma).
\end{eqnarray*}
Hence, for $\hat LG=K^{-1}(LF)$, we have
\begin{eqnarray}
(\hat LG)(\eta)&=&-\sum_{\xi\subset\eta}(-1)^{|\eta\setminus\xi|}\sum_{x\in\xi}
\left(K\left(D_x\star G(\cdot\cup x)\right)\right)(\xi\setminus x)\label{M3}\\
&&+\int_{\R^d}dx\,K^{-1} \left(K\left(B_x\star G(\cdot\cup
x)\right)\right)(\xi).\label{M4}
\end{eqnarray}
A direct application of the definitions of the $K$-transform and $K^{-1}$ 
yields for the sum in (\ref{M3})
\begin{eqnarray*}
&&\sum_{x\in\eta}\sum_{\xi\subset\eta\setminus x}(-1)^{|(\eta\setminus x)\setminus\xi|}\left(K\left(D_x\star G(\cdot\cup x)\right)\right)(\xi)\\
&=&\sum_{x\in\eta}K^{-1}\left(K\left(D_x\star G(\cdot\cup x)\right)\right)(\eta\setminus x)\\
&=&\sum_{x\in\eta}\left(D_x\star G(\cdot\cup x)\right)(\eta\setminus x),
\end{eqnarray*}
and for the integral (\ref{M4})
\[
\int_{\R^d}dx\,K^{-1}\left(K\left(B_x\star G(\cdot\cup x)\right)\right)(\eta)=\int_{\R^d}dx\,\left(B_x\star G(\cdot\cup x)\right)(\eta).
\]

In order to prove the integrability of $\vert\hat LG\vert$ for
$G\in B_{bs}(\Gamma_0)$,
first we note that each $G\in B_{bs}(\Gamma_0)$ can be majorized by
$\vert G\vert\leq C\i_ {\bigsqcup_{n=0}^N\Gamma_\Lambda^{(n)}}$ for some
$C\geq 0$ and for the indicator function
$\i_ {\bigsqcup_{n=0}^N\Gamma_\Lambda^{(n)}}\in B_{bs}(\Gamma_0)$ of some
disjoint union $\bigsqcup_{n=0}^N\Gamma_\Lambda^{(n)}$,
$N\in\N_0,\Lambda\in\mathcal{B}_c(\R^d)$. Hence the proof amounts to show the
integrability of $\vert\hat L\i_ {\bigsqcup_{n=0}^N\Gamma_\Lambda^{(n)}}\vert$
for all $N\in\N$ and all $\Lambda\in\mathcal{B}_c(\R^d)$. This follows from
\begin{eqnarray}
&&\int_{\Gamma_0}d\rho_\mu(\eta)\,\sum_{x\in\eta}\left(\vert D_x\vert\star
\i_{\bigsqcup_{n=0}^N\Gamma_\Lambda^{(n)}}(\cdot\cup x)\right)(\eta\setminus x)\nonumber \\
&&+\int_{\Gamma_0}d\rho_\mu(\eta)\int_{\R^d}dx\,\left(\vert B_x\vert\star
\i_ {\bigsqcup_{n=0}^N\Gamma_\Lambda^{(n)}}(\cdot\cup x)\right)(\eta)\nonumber \\
&\leq&\int_{\Gamma_0}d\rho_\mu(\eta)\,\sum_{x\in\eta}\i_\Lambda(x)
\left(\vert D_x\vert\star
\i_{\bigsqcup_{n=0}^{N-1}\Gamma_\Lambda^{(n)}}\right)(\eta\setminus x)\label{Dima6} \\
&&+\int_\Lambda dx\int_{\Gamma_0}d\rho_\mu(\eta)\,\left(\vert B_x\vert\star
\i_ {\bigsqcup_{n=0}^{N-1}\Gamma_\Lambda^{(n)}}\right)(\eta)\label{Dima7} \\
&=&\int_\Gamma d\mu(\gamma)\,K\left(\sum_{x\in\cdot}\i_\Lambda(x)
\left(\vert D_x\vert\star
\i_{\bigsqcup_{n=0}^{N-1}\Gamma_\Lambda^{(n)}}\right)(\cdot\setminus x)\right)(\gamma)\label{Z3} \\
&&+\int_\Lambda dx\int_\Gamma d\mu(\gamma)\,K\left(\vert B_x\vert\star
\i_ {\bigsqcup_{n=0}^{N-1}\Gamma_\Lambda^{(n)}}\right)(\gamma),\label{Z4}
\end{eqnarray}
where a direct calculation using the definition of the $K$-transform gives for
the integral (\ref{Z3})
\begin{eqnarray*}
&&\int_\Gamma d\mu(\gamma)\,\sum_{x\in\gamma}\i_\Lambda(x)
K\left(\vert D_x\vert\star\i_{\bigsqcup_{n=0}^{N-1}\Gamma_\Lambda^{(n)}}\right)
(\gamma\setminus x)\\
&=&\int_\Gamma d\mu(\gamma)\,\sum_{x\in\gamma_\Lambda}
\left(K\vert D_x\vert\right)(\gamma\setminus x)
\left(K\i_{\bigsqcup_{n=0}^{N-1}\Gamma_\Lambda^{(n)}}\right)(\gamma\setminus x),\end{eqnarray*}
cf.~(\ref{1.5}).

Taking into account that
$\i_{\bigsqcup_{n=0}^{N-1}\Gamma_\Lambda^{(n)}}\in B_{bs}(\Gamma_0)$,
and thus
\[
\left(K\i_{\bigsqcup_{n=0}^{N-1}\Gamma_\Lambda^{(n)}}\right)(\gamma)\leq
(1+\vert\gamma_\Lambda\vert)^{N-1},
\]
one may then bound the sum of the integrals (\ref{Z3}) and (\ref{Z4}) by
\[
\int_\Gamma d\mu(\gamma)\,\vert\gamma_\Lambda\vert^{N-1}\sum_{x\in\gamma_\Lambda}\left(K\vert D_x\vert\right)(\gamma\setminus x)+ \int_\Gamma d\mu(\gamma)\,(1+\vert\gamma_\Lambda\vert)^{N-1}\int_\Lambda dx\,\left(K\vert B_x\vert\right)(\gamma),
\]
which, by (\ref{M5}), shows the required integrability.\hfill$\blacksquare \medskip$

\begin{remark}
\label{Rem200}
Integrability condition (\ref{M5}) is presented for general measures
$\mu\in\mathcal{M}_{\mathrm{fm}}^1(\Gamma)$ and generic birth and death rates
of the type (\ref{Z1}). From the previous proof it is clear that (\ref{M5}) is
the weakest possible integrability condition to state Proposition \ref{Prop1}.
In addition, its proof also shows that for each measure
$\rho\in\mathcal{M}_{\mathrm{lf}}(\Gamma_0)$ such that
$B_x,D_x\in L^1(\Gamma_0,\rho)$ and such that for all $n\in\N_0$ and all
$\Lambda\in\mathcal{B}_c(\R^d)$
\[
\int_{\Gamma_0}d\rho(\eta)\,\left\{\sum_{x\in\eta_\Lambda}
\left(\vert D_x\vert\star
\i_{\Gamma_\Lambda^{(n)}}\right)(\eta\setminus x)
+\int_\Lambda dx\left(\vert B_x\vert\star
\i_{\Gamma_\Lambda^{(n)}}\right)(\eta)\right\}<\infty,
\]
one has $\hat L\left(B_{bs}(\Gamma_0)\right)\subset L^1(\Gamma_0,\rho)$.
Moreover, this integrability condition on
$\rho\in\mathcal{M}_{\mathrm{lf}}(\Gamma_0)$ is the weakest possible
one to yield such an inclusion.
This follows from (\ref{Dima6}), (\ref{Dima7}) and the fact that
$\i_ {\bigsqcup_{n=0}^N\Gamma_\Lambda^{(n)}}=\sum_{n=0}^N\i_{\Gamma_\Lambda^{(n)}}$.
\end{remark}

\begin{remark}
\label{Rem100} Taking into account (\ref{Dima}), we note that:\newline
(1) if each $D_x$ is of the type $D_x=e_\lambda(d_x)$, then the sum in
(\ref{M6}) is given by
\[
\sum_{x\in\eta}\sum_{\xi\subset\eta\setminus x}G(\xi\cup x)
e_\lambda(d_x+1,\xi)e_\lambda(d_x,(\eta\setminus x)\setminus\xi);
\]
(2) Analogously, if $B_x=e_\lambda(b_x)$, then the integral in (\ref{M6}) is
equal to
\[
\sum_{\xi\subset\eta}\int_{\R^d}dx\,G(\xi\cup x)e_\lambda(b_x+1,\xi)
e_\lambda(b_x,\eta\setminus\xi).
\]
\end{remark}

\begin{remark}
\label{Rem2}
For birth and death rates such that
$\vert B_x\vert\leq e_\lambda(b_x), \vert D_x\vert\leq e_\lambda(d_x)$, for
some $0\leq b_x,d_x\in L^1(\R^d,dx)$, and for measures
$\mu\in\mathcal{M}_{\mathrm{fm}}^1(\Gamma)$ that are locally absolutely
continuous with respect to $\pi$ and the correlation function $k_\mu$ fulfills
the so-called Ruelle bound, i.e., $k_\mu\leq e_\lambda(C)$ for some
constant $C>0$, one may replace (\ref{M5}) by the stronger integrability
condition
\begin{equation}
\int_\Lambda dx\,\left(\exp\left(2C\Vert b_x\Vert_{L^1(\R^d,dx)}\right) +
\exp\left(2C\Vert d_x\Vert_{L^1(\R^d,dx)}\right)\right)<\infty,\quad
\forall\,\Lambda\in\mathcal{B}_c(\R^d).\label{P1}
\end{equation}
\end{remark}

\begin{corollary}
\label{Prop2} Let $k:\Gamma_0\to\R^+_0$, $\R^+_0:=\left[0,+\infty\right[$, be 
such that
\begin{equation}
\int_{\Gamma^{(n)}_\Lambda}d\lambda(\eta)\,k(\eta)<\infty\quad
\hbox{for all}\,\,n\in\N_0\,\,
\hbox{and all}\,\,\Lambda \in \mathcal{B}_c(\R^d).\label{Dima8}
\end{equation}
If $B_x,D_x\in L^1(\Gamma_0,k\lambda)$ and for all $n\in\N_0$ and all
$\Lambda\in\mathcal{B}_c(\R^d)$ we have
\[
\int_{\Gamma_0}d\lambda(\eta)\,k(\eta)\left\{\sum_{x\in\eta_\Lambda}
\left(\vert D_x\vert\star
\i_{\Gamma_\Lambda^{(n)}}\right)(\eta\setminus x)
+\int_\Lambda dx\left(\vert B_x\vert\star
\i_{\Gamma_\Lambda^{(n)}}\right)(\eta)\right\}<\infty,
\]
then
\begin{eqnarray}
(\hat L^*k)(\eta)&=&-\int_{\Gamma_0}\!d\lambda(\zeta)\,
k(\zeta\cup\eta)\sum_{x\in \eta}\sum_{\xi\subset\eta\setminus x}
D_x(\zeta\cup\xi)\label{M7} \\
&&+\int_{\Gamma_0}\!d\lambda(\zeta)\,\sum_{x\in \eta}k(\zeta\cup(\eta\setminus x))\sum_{\xi\subset\eta\setminus x} B_x(\zeta\cup\xi),\label{M8}
\end{eqnarray}
for $\lambda$-almost all $\eta\in\Gamma_0$.
\end{corollary}

\noindent
\textbf{Proof.} According to the definition of the dual operator $\hat L^*$,
for all $G\in B_{bs}(\Gamma_0)$ we have
\begin{equation}
\int_{\Gamma_0}d\lambda(\eta)\,(\hat L^*k)(\eta)G(\eta)=
\int_{\Gamma_0}d\lambda(\eta)\,(\hat LG)(\eta) k(\eta).\label{Dima4}
\end{equation}
Due to (\ref{Dima8}), we observe that the measure $k(\eta)\lambda(d\eta)$ on
$\Gamma_0$ is in $\mathcal{M}_{\mathrm{lf}}(\Gamma_0)$. Therefore, according
to Remark \ref{Rem200}, under the fixed assumptions the integral on the
right-hand side of (\ref{Dima4}) is always
finite. The proof then follows by successive applications of Lemmata
\ref{Lmm2} and \ref{Lmm3} to this integral. This procedure applied to the sum
in (\ref{M6}) gives rise to
\begin{eqnarray*}
&&\int_{\Gamma_0}d\lambda(\eta)\,k(\eta)\sum_{x\in\eta}
\left(D_x\star G(\cdot\cup x)\right)(\eta\setminus x)\\
&=&\int_{\R^d}\!dx\int_{\Gamma_0}d\lambda(\eta)\,\left(D_x\star G(\cdot\cup x)\right)(\eta)k(\eta\cup x)\\
&=&\!\int_{\R^d}\!dx\int_{\Gamma_0}\!d\lambda(\eta_1)\!\int_{\Gamma_0}\!d\lambda(\eta_2)D_x(\eta_1\cup\eta_2)
\int_{\Gamma_0}\!d\lambda(\eta_3)\,G(\eta_2\cup\eta_3\cup x)k(\eta_1\cup\eta_2\cup\eta_3\cup x)\\
&=&\!\int_{\Gamma_0}\!d\lambda(\eta_1)\!\int_{\Gamma_0}\!d\lambda(\eta)\,G(\eta)k(\eta_1\cup\eta)\sum_{x\in \eta}\sum_{\xi\subset\eta\setminus x} D_x(\eta_1\cup\xi).
\end{eqnarray*}
Similarly, for the integral expression which appears in (\ref{M6}) we find
\begin{eqnarray*}
&&\int_{\Gamma_0}d\lambda(\eta)\,k(\eta)
\int_{\R^d}dx\,\left(B_x\star G(\cdot\cup x)\right)(\eta)\\
&=&\!\int_{\R^d}\!dx\!\int_{\Gamma_0}\!d\lambda(\eta_1)\!\int_{\Gamma_0}\!d\lambda(\eta_2)\!\int_{\Gamma_0}\!d\lambda(\eta_3)\,B_x(\eta_1\cup\eta_2)G(\eta_2\cup\eta_3\cup x)k(\eta_1\cup\eta_2\cup\eta_3)\\
&=&\!\int_{\Gamma_0}\!d\lambda(\eta_1)\!\int_{\R^d}dx\!\int_{\Gamma_0}\!d\lambda(\eta_2)\!\int_{\Gamma_0}\!d\lambda(\eta_3)\,G(\eta_2\cup\eta_3\cup x)B_x(\eta_1\cup\eta_2)k(\eta_1\cup\eta_2\cup\eta_3)\\
&=&\!\int_{\Gamma_0}\!d\lambda(\eta_1)\!\int_{\Gamma_0}\!d\lambda(\eta)\,G(\eta)\sum_{x\in \eta}\sum_{\xi\subset\eta\setminus x} B_x(\eta_1\cup\xi)k(\eta_1\cup(\eta\setminus x)).
\end{eqnarray*}
Taking into account the density of the space $B_{bs}(\Gamma_0)$ in
$L^1(\Gamma_0,\lambda)$, the required explicit formula
follows.\hfill$\blacksquare \medskip$

\begin{remark}
\label{Rem1}Concerning Corollary \ref{Prop2}, observe that:\newline
(1) if each $D_x$ is of the type $D_x=e_\lambda(d_x)$, then the integral in
(\ref{M7}) is given by
\[
\int_{\Gamma_0}d\lambda(\zeta)k(\eta\cup\zeta)\sum_{x\in\eta}
e_\lambda(d_x+1,\eta\setminus x)e_\lambda(d_x,\zeta);
\]
(2) Analogously, if $B_x=e_\lambda(b_x)$, then (\ref{M8}) is equal to
\[
\int_{\Gamma_0}d\lambda(\zeta)\sum_{x\in\eta}k(\zeta\cup(\eta\setminus x))
e_\lambda(b_x+1,\eta\setminus x)e_\lambda(b_x,\zeta).
\]
\end{remark}

Under quite general assumptions we have derived an explicit form for the
operators $\hat L$, $\hat L^*$ related to the generator of a
birth-and-death dynamics. Within Subsection \ref{Subsection22} framework, this
means that we may describe the underlying dynamics through the time evolution
equations (KE), (QKE), and $\mathrm{(QKE)}^*$, respectively, for observables,
quasi-observables, and correlation functions. The next result concerns a
dynamical description through Bogoliubov functionals.

\begin{proposition}
\label{Prop3} Let $k:\Gamma_0\to\R^+_0$ be such that for all
$\theta\in L^1_\C(\R^d,dx)$ one has
$e_\lambda(\theta)\in L^1_\C(\Gamma_0,k\lambda)$, and the functional
\[
B(\theta) := \int_{\Gamma_0}d\lambda(\eta)\,e_\lambda(\theta,\eta)k(\eta)
\]
is entire on the space $L^1_\C(\R^d,dx)$. If
$B_x,D_x\in L^1(\Gamma_0,k\lambda)$ and
$\hat Le_\lambda(\theta)\in L^1_\C(\Gamma_0,k\lambda)$ for all
$\theta\in L^1_\C(\R^d,dx)$, then
\begin{eqnarray*}
(\tilde LB)(\theta)&=&
-\int_{\Gamma_0}d\lambda(\eta)\,e_\lambda(\theta +1,\eta)
\int_{\R^d}dx\,\theta(x)(D^{\vert\eta\vert+1}B)(\theta,\eta\cup x)D_x(\eta)\\
&&+\int_{\Gamma_0}d\lambda(\eta)\,(D^{\vert\eta\vert}B)(\theta,\eta)
e_\lambda(\theta+1,\eta)\int_{\R^d}dx\,\theta(x)B_x(\eta),
\end{eqnarray*}
for all $\theta\in L^1_\C(\R^d,dx)$.
\end{proposition}

\noindent
\textbf{Proof.} In order to calculate
\[
(\tilde LB)(\theta)=\int_{\Gamma_0}d\lambda(\eta)\,(\hat Le_\lambda(\theta))(\eta)k(\eta),
\]
first we observe that the stated assumptions allow an extension of the
operator $\hat L$ to coherent states $e_\lambda(\theta)$ with
$\theta\in L^1_\C(\R^d,dx)$:
\[
(\hat Le_\lambda(\theta)(\eta)
=-\sum_{x\in\eta}\theta(x)\left(D_x\star e_\lambda(\theta)\right)
(\eta\setminus x)
+\int_{\R^d}dx\,\theta(x)\left(B_x\star e_\lambda(\theta)\right)(\eta).
\]
Using the special simple form (\ref{Dima}) for the $\star$-convolution, a
direct application of Lemma \ref{Lmm2} for $n=2$ yields
\begin{eqnarray*}
&&\int_{\Gamma_0}d\lambda(\eta)\,k(\eta)\sum_{x\in\eta}\theta(x)\left(D_x\star e_\lambda(\theta)\right)(\eta\setminus x)\\
&=&\int_{\R^d}dx\,\theta(x)\int_{\Gamma_0}d\lambda(\eta)\,D_x(\eta)
e_\lambda(\theta+1,\eta)\int_{\Gamma_0}d\lambda(\xi)k(\eta\cup\xi\cup x)
e_\lambda(\theta,\xi).
\end{eqnarray*}
Due to the holomorphicity of $B$ on $L^1_\C(\R^d,dx)$, the latter integral is
equal to $(D^{\vert\eta\cup x\vert}B)(\theta,\eta\cup x)$ cf.~equality
(\ref{Dima1}). Similarly,
\begin{eqnarray*}
&&\int_{\Gamma_0}d\lambda(\eta)\,k(\eta)\int_{\R^d}dx\,\theta(x)\left(B_x\star e_\lambda(\theta)\right)(\eta)\\
&=&\int_{\Gamma_0}d\lambda(\eta)\,(D^{\vert\eta\vert}B)(\theta,\eta)
e_\lambda(\theta+1,\eta)\int_{\R^d}dx\,\theta(x)B_x(\eta).
\end{eqnarray*}
\hfill$\blacksquare \medskip$

\begin{remark}
\label{Rem300}
For functions $k:\Gamma_0\to\R^+_0$ such that $k\leq e_\lambda(C)$ for some
constant $C>0$, the functionals $B$ defined as in Proposition \ref{Prop3} are
well-defined on the whole space $L^1_\C(\R^d,dx)$, cf.~Example \ref{example}.
Moreover, they are entire on $L^1_\C(\R^d,dx)$, see e.g.~\cite{KoKu99c},
\cite{KoKuOl02}. For such functions $k$, one may then state Proposition
\ref{Prop3} just under the assumptions $B_x,D_x\in L^1(\Gamma_0,k\lambda)$ and
$\hat Le_\lambda(\theta)\in L^1_\C(\Gamma_0,k\lambda)$ for all
$\theta\in L^1_\C(\R^d,dx)$.
\end{remark}

\begin{remark}
\label{Rem11}
Proposition \ref{Prop3} is stated for generic birth and death rates of the
type (\ref{Z1}). In applications, the concrete explicit form of such
rates allows a reformulation of Proposition \ref{Prop3}, generally under
much weaker analytical assumptions. For instance, if $B_x$ and $D_x$ are of
the type $B_x=e_\lambda(b_x), D_x=e_\lambda(d)$, where $d$ is independent of
$x$, then the expression for $\tilde LB$ given in Proposition \ref{Prop3}
reduces to
\[
(\tilde LB)(\theta)=
\int_{\R^d}dx\,\theta(x)\left(B(\theta(b_x+1)+b_x)-
\frac{\delta B(\theta(d+1)+d)}{\delta (\theta(d+1)+d)(x)}\right).
\]
In contrast to the general formula, which depends of all variational
derivatives of $B$ at $\theta$, this closed formula only depends on $B$ and
its first variational derivative on a shifted point. Further examples are
presented in Subsection \ref{Subsection32} below. Although in all these
examples Proposition \ref{Prop3} may clearly be stated under much weaker
analytical assumptions, the assumptions in Proposition \ref{Prop3} are
sufficient to state a general result.
\end{remark}

\subsection{Particular models\label{Subsection32}}

Special birth-and-death type models will be presented and discussed
within Subsection \ref{Subsection31} framework. By analogy, all
examples presented are a continuous version of models already known
for lattices systems, see e.g.~\cite{Lig85}, \cite{Lig99}.

\subsubsection{Glauber dynamics\label{Subsubsection321}}

In this birth-and-death type model, particles appear and disappear
according to a death rate identically equal to 1 and to a birth rate
depending on the interaction between particles. More precisely, let
$\phi:\R^d\to\R\cup\{+\infty\}$ be a pair potential, that is, a
Borel measurable function such that $\phi(-x)=\phi(x)\in \R$ for all
$x\in \R^d\setminus\{0\}$, which we assume to be bounded from below,
namely, $\phi \geq -2B_\phi$ on $\R^d$ for some $B_\phi\geq 0$, and which
fulfills the standard integrability condition
\begin{equation}
\int_{\R^d}dx\,\left|e^{-\phi (x)}-1\right|<\infty.\label{P2}
\end{equation}
Given a configuration $\gamma$, the birth rate of a new particle at a site
$x\in\R^d\setminus\gamma$ is then given by $b(x,\gamma)=\exp(-E(x,\gamma))$,
where $E(x,\gamma)$ is a relative energy of interaction between a particle
located at $x$ and the configuration $\gamma$ defined by
\begin{eqnarray}
E(x,\gamma ):=\left\{
\begin{array}{cl}
\displaystyle\sum_{y\in \gamma }\phi (x-y), & \mathrm{if\;}
\displaystyle\sum_{y\in \gamma }|\phi (x-y)|<\infty \\
&  \\
+\infty , & \mathrm{otherwise}
\end{array}
\right. .\label{Dt1}
\end{eqnarray}

In this special example the required conditions (\ref{Z1}) for the birth and
death rates are clearly verified:
\[
d\equiv 1=Ke_\lambda(0),\quad
b(x,\gamma)=e^{-E(x,\gamma)}=\left(Ke_\lambda(e^{-\phi(x-\cdot)}-1)\right)
(\gamma).
\]
Comparing with the general case (Subsection \ref{Subsection31}), the
conditions imposed to the potential $\phi$ lead to a simpler situation. In
fact, the integrability condition (\ref{P2}) implies that for any $C>0$ and
any $\Lambda\in\mathcal{B}_c(\R^d)$ the integral appearing in (\ref{P1}) is
always finite. According to Remark \ref{Rem2}, this implies that for each
measure $\mu\in\mathcal{M}_{\mathrm{fm}}^1(\Gamma)$, locally absolutely
continuous with respect to $\pi$, for which the correlation function fulfills
the Ruelle bound we have $L(\mathcal{FP}(\Gamma))\subset L^1(\Gamma,\mu)$.

The especially simple form of the functions
$B_x=e_\lambda(e^{-\phi(x-\cdot)}-1)$ and $D_x=e_\lambda(0)$ also
allows a simplification of the expressions obtained in Subsection
\ref{Subsection31}. First, as $D_x$ is the unit element of the
$\star$-convolution, using (\ref{Dima}) we obtain for (\ref{M6})
\begin{eqnarray}
(\hat{L}G)(\eta)&=& -\vert \eta \vert G(\eta) + \int_{\R^d}dx\,\left(e_\lambda(e^{-\phi (x - \cdot)}-1)\star G(\cdot \cup x)\right)(\eta)\\
&=&-\vert \eta \vert
G(\eta)+\sum_{\xi\subset\eta}\int_{\R^d}dx\,e^{-E(x,\xi)} G(\xi\cup
x)e_\lambda(e^{-\phi (x - \cdot)}-1,\eta\setminus\xi).\nonumber
\end{eqnarray}
Due to the semi-boundedness of $\phi$, we note that this expression is
well-defined on the whole space $\Gamma_0$. This follows from the fact that
any $G\in B_{bs}(\Gamma_0)$ may be bounded by
$\vert G\vert\leq Ce_\lambda(\i_\Lambda)$, for some $C\geq 0$ and some $\Lambda\in\mathcal{B}_c(\R^d)$, and thus, by (\ref{2.3}),
\begin{eqnarray*}
&&\int_{\R^d}dx\,\left|\left(e_\lambda(e^{-\phi (x - \cdot)}-1)\star
G(\cdot \cup x)\right)(\eta)\right|\\
&\leq&C\int_{\R^d}dx\,\i_\Lambda(\eta)
\left(e_\lambda(\vert e^{-\phi (x - \cdot)}-1\vert)\star
e_\lambda(\i_\Lambda)\right)(\eta)\leq C\left|\Lambda\right|(3+2e^{2B_\phi})^{\vert\eta\vert}.
\end{eqnarray*}
Here $\left|\Lambda\right|$ denotes the volume of the set $\Lambda$. Second,
by Remark \ref{Rem1}, for $\lambda$-almost all $\eta\in\Gamma_0$ we find
\begin{eqnarray}
(\hat L^*k)(\eta)&=&-\int_{\Gamma_0}d\lambda(\zeta)\,
k(\eta\cup\zeta)\sum_{x\in\eta}e_\lambda(1,\eta\setminus x)e_\lambda(0,\zeta)\\
&&+\int_{\Gamma_0}d\lambda(\zeta)\sum_{x\in\eta}k(\zeta\cup(\eta\setminus x))
e_\lambda(e^{-\phi(x-\cdot)},\eta\setminus x)e_\lambda(e^{-\phi(x-\cdot)} -1,\zeta)\nonumber\\
&=&-\vert \eta \vert k(\eta)+\sum_{x\in \eta}e^{-E(x,\eta\setminus
x)} \int_{\Gamma_0}d\lambda(\zeta)\,e_\lambda(e^{-\phi (x -
\cdot)}-1,\zeta) k((\eta\!\setminus x)\cup\zeta).\nonumber
\end{eqnarray}
According to Remark \ref{Rem11}, we also have a simpler form for $\tilde L$,
\begin{equation}
(\tilde LB)(\theta)=
-\int_{\R^d}dx\,\theta(x)\left(\frac{\delta B(\theta)}{\delta\theta(x)}
- B((1+\theta)(e^{-\phi (x -\cdot )}-1)+\theta)\right).
\end{equation}

The Glauber dynamics is the first example which emphasizes the
technical efficacy of our approach to dynamical problems. As a
matter of fact, for a quite general class of pair potentials one may
apply standard Dirichlet forms techniques to $L$ to construct an
equilibrium Glauber dynamics, that is, a Markov process on $\Gamma$
with initial distribution an equilibrium state. This scheme was used
in \cite{LK03} for pair potentials either positive or superstable.
Recently, in \cite{KLRII05}, this construction was extended to a
general case of equilibrium birth-and-death dynamics. However,
starting with a non-equilibrium state, the Dirichlet forms
techniques do not work. Such states can be so far from the
equilibrium ones that one cannot even use the equilibrium Glauber
dynamics (obtained through Dirichlet forms techniques) to construct
the non-equilibrium ones. Within this context, in a recent work
\cite{KoKtZh06} the authors have used the $(\mathrm{QKE})^*$
equation to construct a non-equilibrium Glauber dynamics. That is, a
Markov process on $\Gamma$ starting with a distribution from a wide
class of non-equilibrium initial states, also identified in
\cite{KoKtZh06}. The scheme used is the one described in Subsection
\ref{Subsection22}.

\subsubsection{Linear voter model\label{Subsubsection322}}

Within this model, the individual's motivation to vote is determined
by the attitude of surrounding people towards political participation:
willingness or lack of motivation to vote (perception of voting as a civic
duty or political indifference). Mathematically, this means that, given a
population $\gamma$ of possible voters, an individual $x\in\gamma$ loses his
willingness to vote according to a rate
\[
d(x,\gamma)=\sum_{y\in\gamma} a_-(x,y)=\left(Ka_-(x,\cdot)\right)(\gamma),
\]
for some symmetric function $a_-:\R^d\times\R^d\to\R^+_0$ such that
\[
\sup_{x\in\R^d}\int_{\R^d}dy\,a_-(x,y)<\infty;
\]
while an individual $x$ wins a perception of the importance of joining the
population $\gamma$ according to a rate
\[
b(x,\gamma)=\sum_{y\in\gamma} a_+(x,y)=\left(Ka_+(x,\cdot)\right)(\gamma),
\]
for some symmetric function $a_+:\R^d\times\R^d\to\R^+_0$ such that
\[
\sup_{x\in\R^d}\int_{\R^d}dy\,a_+(x,y)<\infty.
\]

Within Subsection \ref{Subsection31} framework, one straightforwardly derives
from the general case corresponding expressions for this special case:
\begin{eqnarray}
(\hat{L}G)(\eta)&=& -\sum_{x\in\eta}\sum_{y\in\eta\setminus x} a_-(x,y)
\left(G(\eta\setminus y)+G(\eta)\right)\\
&&+ \sum_{y\in\eta}\int_{\R^d}dx\,a_+(x,y)\left(G(\eta\cup
x)+G((\eta\setminus y)\cup x)\right),\nonumber
\end{eqnarray}
and
\begin{eqnarray}
(\hat L^*k)(\eta)&=&
-\int_{\R^d}dy\,k(\eta\cup y)\sum_{x\in\eta}a_-(x,y)-k(\eta)\sum_{x\in\eta}\sum_{y\in\eta\setminus x} a_-(x,y)\\
&&+\int_{\R^d}dy\,\sum_{x\in\eta}k((\eta\setminus x)\cup y)a_+(x,y)
+\sum_{x\in\eta}k(\eta\setminus x)\sum_{y\in\eta\setminus x}
a_+(x,y).\nonumber
\end{eqnarray}
In addition,
\begin{eqnarray}
(\tilde LB)(\theta)&=&\int_{\R^d}dx\int_{\R^d}dy\,a_+(x,y)
(1+\theta(y))\theta(x)\frac{\delta B(\theta)}{\delta\theta(y)}\\
&&-\int_{\R^d}dx\int_{\R^d}dy\,a_-(x,y) (1+\theta(y))\theta(x)
\frac{\delta^2B(\theta)}{\delta\theta(x)\delta\theta(y)}.\nonumber
\end{eqnarray}

\subsubsection{Polynomial voter model\label{Subsubsection323}}

More generally, one may consider rates of polynomial type, that is, the birth and the death rates are of the type
\begin{flalign*}
d(x,\gamma)&=\sum_{\left\{x_1,...,x_q\right\}\subset\gamma}a^{(q)}_x(x_1,...,x_q), &b(x,\gamma)&=\sum_{\left\{x_1,...,x_p\right\}\subset\gamma}a^{(p)}_x(x_1,...,x_p),\\
&=(K\tilde a^{(q)}_x)(\gamma) & &=(K\tilde a^{(p)}_x)(\gamma)
\end{flalign*}
for some symmetric functions $0\leq a^{(q)}_x\in L^1((\R^d)^q,dx_1... dx_q)$,
$0\leq a^{(p)}_x\in L^1((\R^d)^p,dx_1... dx_p)$, $x\in\R^d$, $p,q\in\N$, where
\begin{eqnarray*}
\tilde a^{(i)}_x(\eta):=\left\{
\begin{array}{cl}
a^{(i)}_x(x_1,...,x_i),& \mathrm{if\;}
\eta=\{x_1,...,x_i\}\in\Gamma^{(i)}\\
&  \\
0, & \mathrm{otherwise}
\end{array}
\right. ,\quad i=p,q.
\end{eqnarray*}

A straightforward application of the general results obtained in Subsection
\ref{Subsection31} yields for this case the expressions
\begin{eqnarray}
(\hat LG)(\eta)&=&
-\sum_{x\in\eta}\left(\tilde a^{(q)}_x\star G(\cdot\cup x)\right)
(\eta\setminus x)
+\int_{\R^d}dx\,\left(\tilde a^{(p)}_x\star G(\cdot\cup x)\right)(\eta)\nonumber\\
&=&-\sum_{x\in\eta}\sum_{{\xi\subset\eta\setminus
x}\atop{\vert\xi\vert=q}}
\tilde a^{(q)}_x(\xi)\sum_{\zeta \subset \xi}G(\zeta\cup(\eta\setminus x)\setminus\xi)\nonumber\\
&&+\sum_{{\xi\subset\eta}\atop{\vert\xi\vert=p}}\sum_{\zeta \subset
\xi}\int_{\R^d}dx\,\tilde
a^{(p)}_x(\xi)G(\zeta\cup(\eta\setminus\xi)\cup x)
\end{eqnarray}
and
\begin{eqnarray}
(\hat L^*k)(\eta)&=&-\sum_{i=0}^q\frac{1}{i!}\int_{\Gamma^{(i)}}\!
dm^{(i)}(\zeta)\,k(\zeta\cup\eta)\sum_{x\in \eta}
\sum_{{\xi\subset\eta\setminus x}\atop{\vert\xi\vert=q-i}}
\tilde a^{(q)}_x(\zeta\cup\xi)\\
&&+\sum_{i=0}^p\frac{1}{i!}\int_{\Gamma^{(i)}}\!
dm^{(i)}(\zeta)\,\sum_{x\in \eta}k(\zeta\cup(\eta\setminus x))
\sum_{{\xi\subset\eta\setminus x}\atop{\vert\xi\vert=p-i}} \tilde
a^{(p)}_x(\zeta\cup\xi),\nonumber
\end{eqnarray}
where $m^{(i)}$ is the measure on $\Gamma^{(i)}$ defined in Example
\ref{example} (Subsection \ref{Subsection21}). Moreover,
\begin{eqnarray}
&&(\tilde LB)(\theta)\nonumber\\
&=&-\frac{1}{q!}\int_{\Gamma^{(q)}}dm^{(q)}(\eta)\,e_\lambda(\theta+1,\eta)
\int_{\R^d}dx\,\theta(x)(D^{q+1}B)(\theta,\eta\cup x)
\tilde a^{(q)}_x(\eta)\nonumber\\
&&+\frac{1}{p!}\int_{\Gamma^{(p)}}dm^{(p)}(\eta)\,(D^pB)(\theta,\eta)
e_\lambda(\theta+1,\eta)\int_{\R^d}dx\,\theta(x)\tilde
a^{(p)}_x(\eta).
\end{eqnarray}

\subsubsection{Contact model\label{Subsubsection324}}

The dynamics of a contact model describes the spread of an infectious
disease in a population. Given the set $\gamma$ of infected individuals, an
individual $x\in\gamma$ recovers at a constant rate
$d(x,\gamma)=1=e_\lambda(0)$, while an healthy individual
$x\in\R^d\setminus\gamma$ becomes infected according to an infection spreading
rate which depends on the presence of infected neighbors,
\[
b(x,\gamma)=\lambda\sum_{y\in\gamma} a(x-y)=\left(K(\lambda a(x-\cdot))\right)
(\gamma)
\]
for some function $0\leq a\in L^1(\R^d,dx)$ and some coupling
constant $\lambda\geq 0$. For this particular model, the application
of the general results then yields the following expressions
\begin{equation}
(\hat{L}G)(\eta)=-\vert \eta \vert G(\eta)+
\lambda\sum_{y\in\eta}\int_{\R^d} dx\,a(x-y)\left(G(\eta\cup
x)+G((\eta\setminus y)\cup x)\right),
\end{equation}
and
\begin{eqnarray}
(\hat L^*k)(\eta)&=&-\vert \eta \vert k(\eta)
+\lambda\int_{\R^d}dy\,\sum_{x\in\eta}k((\eta\setminus x)\cup y)a(x-y)\nonumber\\
&&+\lambda\sum_{x\in\eta}k(\eta\setminus x)\sum_{y\in\eta\setminus
x} a(x-y).
\end{eqnarray}

In addition,
\begin{eqnarray}
(\tilde LB)(\theta)&=& -\int_{\R^d}dx\,\theta(x) \frac{\delta
B(\theta)}{\delta\theta(x)}\\&& +
\lambda\int_{\R^d}dy\int_{\R^d}dx\,a(x-y)
(1+\theta(y))\theta(x)\frac{\delta
B(\theta)}{\delta\theta(y)}.\nonumber
\end{eqnarray}
Concerning the corresponding time evolution equation (\ref{Dima3}), the
contact model gives a meaning to the considerations done in Subsection
\ref{Subsection23}. As a matter of fact, one can show that there is a solution
of equation (\ref{Dima3}) only for each finite interval of time. Such a
solution has a radius of analyticity which depends on $t$. For $\lambda\geq 1$ 
the radius of analyticity decreases when $t$ increases \cite{KKP06}. 
Therefore, for $\lambda\geq 1$ equation (\ref{Dima3}) cannot have a
global solution on time.

For finite range functions $0\leq a\in L^1(\R^d,dx)$, $\Vert
a\Vert_{L^1(\R^d,dx)}=1$, being either $a\in L^\infty(\R^d,dx)$ or
$a\in L^{1+\delta}(\R^d,dx)$ for some $\delta >0$, the authors in
\cite{KS06} have proved the existence of a contact process, i.e., a
Markov process on $\Gamma$, starting with an initial configuration
of infected individuals from a wide set of possible initial
configurations. Having in mind that the contact model under
consideration is a continuous version of the well-known contact
model for lattice systems \cite{Lig85}, \cite{Lig99}, the
assumptions in \cite{KS06} are natural. In particular the finite
range assumption, meaning that the infection spreading process only
depends on the influence of infected neighbors on healthy ones.
Concerning the infection spreading rate itself, its additive
character implies that each individual recovers, independently of
the others, after a random exponentially distributed time
\cite{KS06}. Within Subsection \ref{Subsection22} framework, in a
recent work \cite{KKP06} the authors have used the
$(\mathrm{QKE})^*$ equation to extend the previous existence result
to Markov processes on $\Gamma$ starting with an initial
distribution. Besides the construction of the processes, the scheme
used allows to identify all invariant measures for such contact
processes.

\section{Conservative dynamics\label{Section4}}

In contrast to the birth-and-death dynamics, in the following dynamics
there is conservation on the number of particles involved.

\subsection{Hopping particles: the general case\label{subsection41}}

Dynamically, in a hopping particles system, at each random moment of time
particles randomly hop from one site to another according to a rate depending
on the configuration of the whole system at that time. In terms of generators
this behaviour is informally described by
\begin{equation}
(LF)(\gamma)= \sum_{x\in\gamma}\int_{\R^d}dy\,c(x,y,\gamma)\left(F(\gamma\setminus x\cup y)-F(\gamma)\right),\label{W1}
\end{equation}
where the coefficient $c(x,y,\gamma)\geq 0$ indicates the rate at which a
particle located at $x$ in a configuration $\gamma$ hops to a site $y$.

To give a rigorous meaning to the right-hand side of (\ref{W1}), we shall
consider measures $\mu\in\mathcal{M}_{\mathrm{fm}}^1(\Gamma)$ such that
$c(x,y,\cdot)\in L^1(\Gamma,\mu)$, $x,y\in\R^d$ and, for
all $n\in\N_0$ and all $\Lambda\in\mathcal{B}_c(\R^d)$ which fulfills the
integrability condition
\begin{equation}
\int_\Gamma d\mu(\gamma)\,|\gamma _\Lambda |^n
\sum_{x\in\gamma}\int_{\R^d}dy\, c(x,y,\gamma)\left(\i_\Lambda(x)+\i_\Lambda(y)\right)<\infty.\label{Wete2}
\end{equation}
In this way, given a cylinder function $F\in\mathcal{FP}(\Gamma)$, $\left|F(\gamma)\right|=\left|F(\gamma_\Lambda)\right|\leq C(1+\vert\gamma_\Lambda\vert)^N$
for some $\Lambda\in \mathcal{B}_c(\R^d), N\in\N_0, C\geq 0$, for all
$\gamma\in\Gamma$ one finds
\[
\vert F(\gamma\setminus x\cup y)-F(\gamma)\vert\leq
2C(2+\vert\gamma_\Lambda\vert)^N(\i_\Lambda(x)+\i_\Lambda(y)).
\]
By (\ref{Wete2}), this implies that $\mu$-a.e.~the right-hand side of
(\ref{W1}) is well-defined and finite and, moreover, it defines an element in
$L^1(\Gamma,\mu)$.

Given a family of functions $C_{x,y}:\Gamma_0\to\R$, $x,y\in\R^d$, such that
$KC_{x,y}\geq 0$, in the following we wish to consider the case
\begin{equation}
c(x,y,\gamma)=(KC_{x,y})(\gamma\setminus x).\label{Dima10}
\end{equation}
Therefore, we shall restrict the previous class of measures in
$\mathcal{M}_{\mathrm{fm}}^1(\Gamma)$ to all measures
$\mu\in\mathcal{M}_{\mathrm{fm}}^1(\Gamma)$ such that
$C_{x,y}\in L^1(\Gamma_0,\rho_\mu)$, $x,y\in\R^d$, and
\begin{equation}
\int_\Gamma d\mu(\gamma)\,|\gamma _\Lambda |^n
\sum_{x\in\gamma}\int_{\R^d}dy\, \left(K\vert C_{x,y}\vert\right)(\gamma\!\setminus\!x)\left(\i_\Lambda(x)+\i_\Lambda(y)\right)<\infty\label{W2}
\end{equation}
for all $n\in\N_0$ and all $\Lambda\in\mathcal{B}_c(\R^d)$. In this way, the
$K$-transform of each $C_{x,y}$, $x,y\in\R^d$, is well-defined,
$KC_{x,y}\in L^1(\Gamma,\mu)$, and
$L(\mathcal{FP}(\Gamma))\subset L^1(\Gamma,\mu)$.

\begin{proposition}
\label{Prop6} The action of the operator $\hat L$ on functions
$G\in B_{bs}(\Gamma_0)$ is given by
\[
(\hat LG)(\eta)=\sum_{x\in\eta}\int_{\R^d}dy\,\left(C_{x,y}\star \left(G(\cdot\cup y)-G(\cdot\cup x)\right)\right)(\eta\setminus x),
\]
for $\rho_\mu$-almost all $\eta\in\Gamma_0$. We have
$\hat L\left(B_{bs}(\Gamma_0)\right)\subset L^1(\Gamma_0,\rho_\mu)$.
\end{proposition}

\noindent
\textbf{Proof.} By the definition of the space $\mathcal{FP}(\Gamma)$, any
element $F\in\mathcal{FP}(\Gamma)$ is of the form $F=KG$ for some
$G\in B_{bs}(\Gamma_0)$. The properties of the $K$-transform, namely, its
algebraic action (\ref{1.5}), then allow to rewrite $LF$ as
\begin{eqnarray*}
(LF)(\gamma)&=&\sum_{x\in\gamma}\int_{\{y:y\notin\gamma\setminus x\}}dy\,
c(x,y,\gamma)
\left(K\left(G(\cdot\cup y)-G(\cdot\cup x)\right)\right)
(\gamma\!\setminus\!x)\\
&=&\sum_{x\in\gamma}\int_{\R^d}dy\,
\left(K\left(C_{x,y}\star (G(\cdot\cup y)-G(\cdot\cup x))\right)\right)(\gamma\!\setminus\!x).
\end{eqnarray*}
Hence
\begin{eqnarray*}
(\hat LG)(\eta)&=&
K^{-1}\left(\sum_{x\in\cdot}\int_{\R^d}dy\,
\left(K\left(C_{x,y}\star (G(\cdot\cup y)-G(\cdot\cup x))\right)\right)(\cdot\setminus x)\right)(\eta)\\
&=&\sum_{\xi\subset\eta}(-1)^{\vert\eta\setminus\xi\vert}
\sum_{x\in\xi}\int_{\R^d}dy\,
\left(K\left(C_{x,y}\star (G(\cdot\cup y)-G(\cdot\cup x))\right)\right)
(\xi\!\setminus\!x)\\
&=&\int_{\R^d}dy\,\sum_{\xi\subset\eta}(-1)^{\vert\eta\setminus\xi\vert}\sum_{x\in\xi}\left(K\left(C_{x,y}\star (G(\cdot\cup y)-G(\cdot\cup x))\right)\right)(\xi\!\setminus\!x)\\
&=&\int_{\R^d}dy\,\sum_{x\in\eta}\sum_{\xi\subset\eta\setminus x}(-1)^{\vert(\eta\setminus x)\setminus\xi\vert}\left(K\left(C_{x,y}\star (G(\cdot\cup y)-G(\cdot\cup x))\right)\right)(\xi)\\
&=&\sum_{x\in\eta}\int_{\R^d}dy\,\left(C_{x,y}\star \left(G(\cdot\cup y)-G(\cdot\cup x)\right)\right)(\eta\!\setminus\!x).
\end{eqnarray*}

As in the proof of Proposition \ref{Prop1}, to check the required inclusion
amounts to prove that for all $N\in\N$ and all $\Lambda\in\mathcal{B}_c(\R^d)$
one has
$\hat L\i_ {\bigsqcup_{n=0}^N\Gamma_\Lambda^{(n)}}\in L^1(\Gamma_0,\rho_\mu)$.
Similar arguments then yield
\begin{eqnarray*}
&&\int_{\Gamma_0}d\rho_\mu(\eta)\,\left|\left(\hat L\i_ {\bigsqcup_{n=0}^N\Gamma_\Lambda^{(n)}}\right)(\eta)\right|\\
&\leq&\int_{\Gamma_0}d\rho_\mu(\eta)\,\sum_{x\in\eta}\int_{\R^d}dy\,\left(\vert C_{x,y}\vert\star
\i_{\bigsqcup_{n=0}^N\Gamma_\Lambda^{(n)}}(\cdot\cup y)\right)(\eta\!\setminus\!x)\\
&&+\int_{\Gamma_0}d\rho_\mu(\eta)\,\sum_{x\in\eta}\int_{\R^d}dy\,\left(\vert C_{x,y}\vert\star
\i_{\bigsqcup_{n=0}^N\Gamma_\Lambda^{(n)}}(\cdot\cup x)\right)(\eta\!\setminus\!x)\\
&\leq&\int_\Lambda dy\,\int_\Gamma d\mu(\gamma)\,(1+\vert\gamma_\Lambda\vert)^{N-1}\sum_{x\in\gamma}\left(K\vert C_{x,y}\vert\right)(\gamma\!\setminus\!x)\\
&&+\int_{\R^d} dy\,\int_\Gamma d\mu(\gamma)\,\vert\gamma_\Lambda\vert^{N-1}\sum_{x\in\gamma_\Lambda}\left(K\vert C_{x,y}\vert\right)(\gamma\!\setminus\!x),
\end{eqnarray*}
which, by (\ref{W2}), complete the proof.\hfill$\blacksquare \medskip$

\begin{remark}Similarly to the proof of Proposition \ref{Prop1}, the proof of
Proposition \ref{Prop6} shows that (\ref{W2}) is the weakest possible
integrability condition to state Proposition \ref{Prop6} for generic measures
$\mu\in\mathcal{M}_{\mathrm{fm}}^1(\Gamma)$ and generic rates $c$ of the type
(\ref{Dima10}). Its proof also shows that for each measure
$\rho\in\mathcal{M}_{\mathrm{lf}}(\Gamma_0)$ such that
$C_{x,y}\in L^1(\Gamma_0,\rho)$ and such that for all $n\in\N_0$ and all
$\Lambda\in\mathcal{B}_c(\R^d)$
\[
\int_{\Gamma_0}d\rho(\eta)\,\sum_{x\in\eta}\int_{\R^d}dy\,\left(\vert C_{x,y}\vert\star\i_{\Gamma_\Lambda^{(n)}}\right)(\eta\!\setminus\!x)
\left(\i_\Lambda(x)+\i_\Lambda(y)\right)<\infty,
\]
we have $\hat L\left(B_{bs}(\Gamma_0)\right)\subset L^1(\Gamma_0,\rho)$. This
integrability condition for measures
$\rho\in\mathcal{M}_{\mathrm{lf}}(\Gamma_0)$ is the weakest possible one to
yield this inclusion.
\end{remark}

\begin{remark}
Concerning Proposition \ref{Prop6} we note that if each $C_{x,y}$ is of the
type $C_{x,y}=e_\lambda(c_{x,y})$, then
\[
(\hat LG)(\eta)=\sum_{x\in\eta}\sum_{\xi\subset\eta\setminus x}\int_{\R^d}dy\,(G(\xi\cup y)-G(\xi\cup x))e_\lambda(c_{x,y}+1,\xi)e_\lambda(c_{x,y},(\eta\setminus x)\setminus\xi),
\]
cf.~equality (\ref{Dima}).
\end{remark}

\begin{remark}
For rates $C_{x,y}$ such that $\vert C_{x,y}\vert\leq e_\lambda(c_{x,y})$ for
some $0\leq c_{x,y}\in L^1(\R^d,dx)$, and for measures
$\mu\in\mathcal{M}_{\mathrm{fm}}^1(\Gamma)$ that are locally absolutely
continuous with respect to $\pi$ and the correlation function $k_\mu$ fulfills
the Ruelle bound for some constant $C>0$, one may replace (\ref{W2}) by the
stronger integrability condition
\[
\int_{\R^d}dx\int_{\R^d}dy\,\exp(2C\Vert c_{x,y}\Vert_{L^1(\R^d,dx)})\left(\i_\Lambda(x)+\i_\Lambda(y)\right)<\infty,\quad \forall\,\Lambda\in\mathcal{B}_c(\R^d).
\]
\end{remark}

Similarly to the proof of Corollary \ref{Prop2}, successive applications of
Lemmata \ref{Lmm2} and \ref{Lmm3} lead to the next result.
\begin{proposition}
\label{Prop7} Let $k:\Gamma_0\to\R^+_0$ be such that
\[
\int_{\Gamma^{(n)}_\Lambda}d\lambda(\eta)\,k(\eta)<\infty\quad
\hbox{for all}\,\,n\in\N_0\,\,
\hbox{and all}\,\,\Lambda \in \mathcal{B}_c(\R^d).
\]
If $C_{x,y}\in L^1(\Gamma_0,k\lambda)$ and for all $n\in\N_0$ and all
$\Lambda\in\mathcal{B}_c(\R^d)$ we have
\[
\int_{\Gamma_0}d\lambda(\eta)\,k(\eta)\sum_{x\in\eta}\int_{\R^d}dy\,\left(\vert C_{x,y}\vert\star\i_{\Gamma_\Lambda^{(n)}}\right)(\eta\!\setminus\!x)
\left(\i_\Lambda(x)+\i_\Lambda(y)\right)<\infty,
\]
then the action of the operator $\hat L^*$ on $k$ is given by
\begin{eqnarray*}
(\hat L^*k)(\eta)&=&\sum_{y\in\eta}\int_{\R^d}dx
\int_{\Gamma_0}d\lambda(\xi)\,k(\xi\cup(\eta\!\setminus\!y)\cup x)
\sum_{\zeta\subset\eta\setminus y}C_{x,y}(\xi\cup\zeta)\\
&&-\int_{\Gamma_0}d\lambda(\xi)\,k(\xi\cup\eta)\sum_{x\in\eta}\sum_{\zeta\subset\eta\setminus x}\int_{\R^d}dy\,C_{x,y}(\xi\cup\zeta),
\end{eqnarray*}
for $\lambda$-almost all $\eta\in\Gamma_0$.
\end{proposition}

\begin{remark}
\label{Rem3} Under the conditions of Proposition \ref{Prop7}, if each
$C_{x,y}$ is of the type $C_{x,y}=e_\lambda(c_{x,y})$, then
\begin{eqnarray*}
(\hat L^*k)(\eta)&=&\sum_{y\in\eta}\int_{\R^d}dx\,
e_\lambda(c_{x,y}+1,\eta\!\setminus\!y)\int_{\Gamma_0}d\lambda(\xi)\,k(\xi\cup(\eta\!\setminus\!y)\cup x)e_\lambda(c_{x,y},\xi)\\
&&-\int_{\Gamma_0}d\lambda(\xi)\,k(\xi\cup\eta)\sum_{x\in\eta}
\int_{\R^d}dy\,e_\lambda(c_{x,y}+1,\eta\!\setminus\!x)e_\lambda(c_{x,y},\xi).
\end{eqnarray*}
\end{remark}

\begin{proposition}
\label{Prop8} Let $k:\Gamma_0\to\R^+_0$ be such that
$e_\lambda(\theta)\in L^1_\C(\Gamma_0,k\lambda)$ for all
$\theta\in L^1_\C(\R^d,dx)$, and the functional
\[
B(\theta) := \int_{\Gamma_0}d\lambda(\eta)\,e_\lambda(\theta,\eta)k(\eta)
\]
is entire on the space $L^1_\C(\R^d,dx)$. If
$C_{x,y}\in L^1(\Gamma_0,k\lambda)$ and
$\hat Le_\lambda(\theta)\in L^1_\C(\Gamma_0,k\lambda)$ for all
$\theta\in L^1_\C(\R^d,dx)$, then for all $\theta\in L^1_\C(\R^d,dx)$ we have
\[
(\tilde LB)(\theta)=
\int_{\Gamma_0}d\lambda(\eta)e_\lambda(\theta+1,\eta)\int_{\R^d}dx\,
(D^{\vert\eta\vert+1}B)(\theta,\eta\cup x)\int_{\R^d}dy\,
(\theta(y)-\theta(x))C_{x,y}(\eta).
\]
\end{proposition}

\noindent
\textbf{Proof.} This proof follows similarly to the proof of Proposition
\ref{Prop3}. In this case we obtain
\begin{eqnarray*}
(\hat Le_\lambda(\theta))(\eta)&=&\sum_{x\in\eta}\int_{\R^d}dy\,(\theta(y)-
\theta(x))(C_{x,y}\star e_\lambda(\theta))(\eta\!\setminus\!x)\\
&=&\sum_{x\in\eta}\int_{\R^d}dy\,(\theta(y)-
\theta(x))\sum_{\xi \subset \eta\setminus x}C_{x,y}(\xi)
e_{\lambda }\left(\theta+1,\xi \right)e_\lambda(\theta,(\eta\setminus x)\setminus\xi),
\end{eqnarray*}
where we have used the expression (\ref{Dima}) concerning the
$\star$-convolution. Arguments similar to those used in the proof of
Proposition \ref{Prop3} lead then to
\begin{eqnarray*}
&&\int_{\Gamma_0}d\lambda(\eta)\,k(\eta)\,(\hat Le_\lambda(\theta))(\eta)\\
&=&\int_{\R^d}dx\int_{\Gamma_0}d\lambda(\eta)\,(D^{\vert\eta\cup x\vert}B)(\theta,\eta\cup x)e_\lambda(\theta+1,\eta)\int_{\R^d}dy\,
(\theta(y)-\theta(x))C_{x,y}(\eta).\\
\end{eqnarray*}
\hfill$\blacksquare \medskip$

\begin{remark}
According to Remark \ref{Rem300}, for functions $k:\Gamma_0\to\R^+_0$ such
that $k\leq e_\lambda(C)$ for some constant $C>0$, one may state Proposition
\ref{Prop8} just under the assumptions $C_{x,y}\in L^1(\Gamma_0,k\lambda)$ and
$\hat Le_\lambda(\theta)\in L^1_\C(\Gamma_0,k\lambda)$ for all
$\theta\in L^1_\C(\R^d,dx)$.
\end{remark}

\begin{remark}
\label{Rem12}
As before, in applications, the concrete explicit form of the rate $C_{x,y}$
allows a reformulation of Proposition \ref{Prop8}, in general under much
weaker analytical assumptions. For instance, if $C_{x,y}=e_\lambda(c_y)$ for
some function $c_y$ which is independent of $x$, then the expression for
$\tilde LB$ given in Proposition \ref{Prop8} reduces to
\[
(\tilde LB)(\theta)=
\int_{\R^d}dx\int_{\R^d}dy\,(\theta(y)-\theta(x))
\frac{\delta B(\theta(c_y+1)+c_y)}{\delta (\theta(c_y+1)+c_y)(x)}.
\]
In contrast to the general formula, which depends of all variational
derivatives of $B$ at $\theta$, this closed formula only depends on the first
variational derivative of $B$ on a shifted point. Further examples are
presented in Subsection \ref{Subsection42}. Although in all such
examples Proposition \ref{Prop8} may clearly be stated under much weaker
analytical assumptions, the assumptions in Proposition \ref{Prop8} are
sufficient to state a general result.
\end{remark}

\subsection{Particular models\label{Subsection42}}

Special hopping particles models will be presented and discussed within
Subsection \ref{subsection41} framework. By analogy, such examples are a
continuous version of models already known for lattice systems.

\subsubsection{Kawasaki dynamics\label{Subsubsection421}}

In such a dynamics particles hop over the space $\R^d$ according to a rate
which depends on the interaction between particles. This means that given a
pair potential $\phi:\R^d\to \R\cup\{+\infty\}$, the rate $c$ is of the form
\begin{eqnarray}
c(x,y,\gamma)\!&=&\!c_s(x,y,\gamma)=a(x-y)
e^{sE(x,\gamma\setminus x)-(1-s)E(y,\gamma)}\nonumber \\
\!&=&\!K\!\left(a(x-y)e^{(s-1)\phi(x-y)}e_\lambda(e^{s\phi(x-\cdot)-(1-s)\phi(y-\cdot)}-1)\right)(\gamma\!\setminus\!x)\label{Dt2}
\end{eqnarray}
for some $s\in\left[0,1\right]$. Here $a:\R^d\to\R^+_0$ and $E$ is a relative
energy defined as in (\ref{Dt1}).

For $a\in L^1(\R^d,dx)$ and for $\phi$ bounded from below and
fulfilling the integrability condition (\ref{P2}), the condition
(\ref{W2}) is always fulfilled, for instance, by any Gibbs measure
$\mu\in\mathcal{M}_{\mathrm{fm}}^1(\Gamma)$ corresponding to $\phi$
for which the correlation function fulfills the Ruelle bound. We
recall that a probability measure $\mu$ on $(\Gamma,
\mathcal{B}(\Gamma))$ is called a Gibbs or an equilibrium measure if
it fulfills the integral equation
\begin{equation}
\int_\Gamma d\mu (\gamma)\,\sum_{x\in \gamma}H(x,\gamma\!\setminus\!x)
=\int_\Gamma d\mu (\gamma)\int_{\R^d}dx\,H(x,\gamma )e^{-E(x,\gamma )}\label{1.3}
\end{equation}
for all positive measurable functions $H:\R^d\times\Gamma\to \R$ (\cite[Theorem 2]{NZ79}, see also \cite[Theorem 3.12]{KoKu98}, \cite[Appendix A.1]{K00}).
Correlation measures corresponding to such a class of measures are always
absolutely
continuous with respect to the Lebesgue-Poisson measure $\lambda$. For Gibbs
measures described as before, the integrability condition (\ref{W2}) follows
as a consequence of (\ref{1.3}), applying the assumptions on $\phi$ and the
Ruelle boundedness. For such Gibbs measures $\mu$ and for $a$ being, in
addition, an even function, it is shown in \cite{KLRII05} the existence
of an equilibrium Kawasaki dynamics, i.e., a Markov process on $\Gamma$ which
generator is given by (\ref{W1}) for $c$ defined as in (\ref{Dt2}). Such a
process has $\mu$ as an invariant measure.

The general results obtained in Subsection \ref{subsection41} yield
for the Kawasaki dynamics the expressions
\begin{eqnarray}
&&(\hat LG)(\eta)=\\
&&\sum_{x\in\eta}\sum_{\xi\subset\eta\setminus
x}e^{sE(x,\xi)}\int_{\R^d}dy\, a(x-y)e^{(s-1)E(y,\xi\cup x)}\nonumber\\
&&\cdot
e_\lambda(e^{s\phi(x-\cdot)-(1-s)\phi(y-\cdot)}-1,(\eta\setminus
x)\setminus\xi)(G(\xi\cup y)-G(\xi\cup x)),\nonumber
\end{eqnarray}
and
\begin{eqnarray}
&&(\hat L^*k)(\eta)\\
&=&\sum_{y\in\eta}\int_{\R^d}dx\,a(x-y)e^{sE(x,\eta\setminus y)-(1-s)E(y,\eta\setminus y\cup x)}\nonumber\\
&&\cdot\int_{\Gamma_0}d\lambda(\xi)\,k(\xi\cup(\eta\!\setminus\!y)\cup
x)
e_\lambda(e^{s\phi(x-\cdot)-(1-s)\phi(y-\cdot)}-1,\xi)\nonumber\\
&&-\int_{\Gamma_0}d\lambda(\xi)\,k(\xi\cup\eta)\nonumber\\
&&\cdot\sum_{x\in\eta}\int_{\R^d}dy\,a(x-y)e^{sE(x,\eta\setminus
x)-(1-s)E(y,\eta)}e_\lambda(e^{s\phi(x-\cdot)-(1-s)\phi(y-\cdot)}-1,\xi),
\nonumber
\end{eqnarray}
where we have taken into account Remark \ref{Rem3}. In terms of Bogoliubov
functionals, Proposition \ref{Prop8} leads to
\begin{eqnarray}
&&(\tilde LB)(\theta)\\
&=&\int_{\Gamma_0}d\lambda(\eta)e_\lambda(\theta+1,\eta)\int_{\R^d}dx\,
(D^{\vert\eta\vert+1}B)(\theta,\eta\cup x)\nonumber\\
&&\cdot\int_{\R^d}dy\,
a(x-y)e^{(s-1)\phi(x-y)}(\theta(y)-\theta(x))e_\lambda(e^{s\phi(x-\cdot)-(1-s)\phi(y-\cdot)}-1,\eta).
\nonumber
\end{eqnarray}
In particular, for $s=0$, one obtains
\begin{eqnarray}
&&(\tilde LB)(\theta)\\&=&
\int_{\R^d}dx\int_{\R^d}dy\,a(x-y)e^{-\phi(x-y)}(\theta(y)-\theta(x))
\frac{\delta B((1+\theta)(e^{-\phi(y-\cdot)}-1)+\theta)} {\delta
((1+\theta)(e^{-\phi(y-\cdot)}-1)+\theta)(x)},\nonumber
\end{eqnarray}
cf.~Remark \ref{Rem12}.

\begin{remark}
In the case $s=0$, in a recent work \cite{FKL06} the authors have shown that
in the high-temperature-low activity regime the scaling limit (of a Kac type)
of an equilibrium Kawasaki dynamics yields in the limit an equilibrium
Glauber dynamics. More precisely, given an even function
$0\leq a\in L^1(\R^d,dx)$ and a stable pair potential $\phi$, i.e.,
\[
\exists B_\phi\geq 0: \sum_{\{x,y\}\subset\eta}\phi(x-y)\geq -B_\phi|\eta|,\ \forall\,\eta\in\Gamma_0,
\]
such that
\[
\int_{\R^d}dx\,\left|e^{-\phi (x)}-1\right|<\left(2e^{1+2B_\phi}\right)^{-1}
\]
(high temperature-high temperature regime), the authors have considered an
equilibrium Kawasaki dynamics which generator $L_\varepsilon$ is given by
(\ref{W1}) for $c$ defined as in (\ref{Dt2}) for $s=0$ and $a$ replaced by the
function $\varepsilon^da(\varepsilon\cdot)$. We observe that such a dynamics
exists due to \cite{KLRII05}. Then it has been shown that the generators
$L_\varepsilon$ converge to
\[
-\alpha\sum_{x\in\gamma}\left(F(\gamma\setminus x)-F(\gamma)\right)
-\alpha\int_{\R^d}dx\,e^{-E(x,\gamma)}\left(F(\gamma\cup x)-F(\gamma)\right),
\]
which is the generator of an equilibrium Glauber dynamics. Here
$\alpha:=k_\mu^{(1)}\int_{\R^d}dx\,a(x)$ for
$k_\mu^{(1)}:=k_\mu\!\!\upharpoonright_{\Gamma^{(1)}}$ being the first
correlation function of the initial distribution $\mu$.
\end{remark}

\subsubsection{Free hopping particles\label{Subsubsection422}}

In the free Kawasaki
dynamics case one has $\phi\equiv 0$, meaning that particles hop freely over
the space $\R^d$. Therefore, all previous considerations hold for this special
case. In particular, for every even function $0\leq a\in L^1(\R^d,dx)$ the
construction done in \cite{KLRII05} yields the existence of an equilibrium
free Kawasaki dynamics. Actually, in this case the generator $L$ is a second
quantization operator which leads to a simpler situation. The existence result
extends to the non-equilibrium case \cite{KLR05} for a wide class of initial
configurations also identified in \cite{KLR05}. This allows the study done in
\cite{KoKuSiSt05} of the large time asymptotic behaviours and hydrodynamical
limits.

\subsubsection{Polynomial rates\label{Subsubsection423}}

In applications one may also consider rates of polynomial type, i.e.,
\[
c(x,y,\gamma)=\sum_{\left\{x_1,...,x_p\right\}\subset\gamma\setminus x}
c^{(p)}_{x,y}(x_1,...,x_p)=(K\tilde c^{(p)}_{x,y})(\gamma\setminus x)
\]
for some symmetric function $0\leq c^{(p)}_{x,y}\in L^1((\R^d)^p,dx_1... dx_p)$, $x\in\R^d$, $p\in\N$, where
\begin{eqnarray*}
(\tilde c^{(p)}_{x,y})(\eta):=\left\{
\begin{array}{cl}
c^{(p)}_{x,y}(x_1,...,x_p),& \mathrm{if\;}
\eta=\{x_1,...,x_p\}\in\Gamma^{(p)}\\
&  \\
0, & \mathrm{otherwise}
\end{array}
\right. .
\end{eqnarray*}
A straightforward application of the general results obtained in Subsection
\ref{subsection41} yields for this case the expressions
\begin{equation}
(\hat LG)(\eta)=\sum_{x\in\eta}\int_{\R^d}dy\,\left(\tilde c^{(p)}_{x,y}\star \left(G(\cdot\cup y)-G(\cdot\cup x)\right)\right)(\eta\setminus x),
\end{equation}
and
\begin{eqnarray}
&&(\hat L^*k)(\eta)
\nonumber\\&=&\sum_{y\in\eta}\sum_{i=0}^p\frac{1}{i!}
\int_{\Gamma^{(i)}}dm^{(i)}(\xi)\int_{\R^d}dx\,k(\xi\cup(\eta\!\setminus\!y)\cup
x)
\sum_{{\zeta\subset\eta\setminus y}\atop{\vert\zeta\vert=p-i}}\tilde c^{(p)}_{x,y}(\xi\cup\zeta) \nonumber\\
&&-\sum_{i=0}^p\frac{1}{i!}\int_{\Gamma^{(i)}}dm^{(i)}(\xi)\,k(\xi\cup\eta)
\sum_{x\in\eta}\sum_{{\zeta\subset\eta\setminus
x}\atop{\vert\zeta\vert=p-i}}\int_{\R^d}dy\,\tilde
c^{(p)}_{x,y}(\xi\cup\zeta),
\end{eqnarray}
where $m^{(i)}$ is the measure on $\Gamma^{(i)}$ defined in Example
\ref{example} (Subsection \ref{Subsection21}). In terms of Bogoliubov
functionals, the statement of Proposition \ref{Prop8} leads now to
\begin{eqnarray}
&&(\tilde LB)(\theta)=\\
&&\frac{1}{p!}\int_{\Gamma^{(p)}}dm^{(p)}(\eta)\,e_\lambda(\theta+1,\eta)
\int_{\R^d}dx\,(D^{p+1}B)(\theta,\eta\cup x)\int_{\R^d}dy\, \tilde
c^{(p)}_{x,y}(\eta)(\theta(y)-\theta(x)).\nonumber
\end{eqnarray}

As a particular realization, one may consider
\[
c(x,y,\gamma)=b(x,y)+\sum_{x_1\in\gamma\setminus x}c^{(1)}_{x,y}(x_1)
=K\left(b(x,y)e_\lambda(0)+\tilde c^{(1)}_{x,y}\right)(\gamma\setminus x),
\]
where $b$ is a function independent of $\gamma$. From the previous
considerations we obtain
\begin{eqnarray}
(\hat LG)(\eta)&=&\sum_{x\in\eta}\int_{\R^d}dy\, b(x,y)(G((\eta\setminus x)\cup y)- G(\eta))\\
&&+\sum_{x\in\eta}\sum_{x_1\in\eta\setminus x}\int_{\R^d}dy\,c^{(1)}_{x,y}(x_1)
(G((\eta\setminus \{x,x_1\})\cup y)- G(\eta\setminus x_1))\nonumber\\
&&+\sum_{x\in\eta}\int_{\R^d}dy\,(G((\eta\setminus x)\cup y)-
G(\eta))\sum_{x_1\in\eta\setminus x}c^{(1)}_{x,y}(x_1),\nonumber
\end{eqnarray}
and
\begin{eqnarray}
(\hat L^*k)(\eta)&=&\sum_{y\in\eta}\int_{\R^d}dx_1\int_{\R^d}dx\,k(x_1\cup(\eta\setminus y)\cup x)c^{(1)}_{x,y}(x_1)\\
&&-\int_{\R^d}dx_1\,k(\eta\cup x_1)\sum_{x\in\eta}\int_{\R^d}dy\,c^{(1)}_{x,y}(x_1)\nonumber\\
&&+\sum_{y\in\eta}\int_{\R^d}dx\,k((\eta\setminus y)\cup
x)\Big(b(x,y)+
\sum_{x_1\in\eta\setminus y}c^{(1)}_{x,y}(x_1)\Big)\nonumber\\
&&-k(\eta)\sum_{x\in\eta}\int_{\R^d}dy\,\Big(b(x,y)+\sum_{x_1\in\eta\setminus
x}c^{(1)}_{x,y}(x_1)\Big).\nonumber
\end{eqnarray}
In addition,
\begin{eqnarray}
(\tilde LB)(\theta)&=&
\int_{\R^d}dx\,\frac{\delta B(\theta)}{\delta \theta(x)}\int_{\R^d}dy\,b(x,y)
(\theta(y)-\theta(x))\\
&&+\int_{\R^d}dx_1\,(\theta(x_1)+1)\int_{\R^d}dx\,\frac{\delta^2B(\theta)}{\delta
\theta(x_1)\delta\theta(x)}\int_{\R^d}dy\,c^{(1)}_{x,y}(x_1)(\theta(y)-\theta(x)).\nonumber
\end{eqnarray}

\subsection{Other conservative jumps processes}\label{Subsection44}

Before we have analyzed individual hops of particles. We may also
analyze hops of groups of $n\geq 2$ particles. Dynamically this
means that at each random moment of time a group of $n$ particles
randomly hops over the space $\R^d$ according to a rate which
depends on the configuration of the whole system at that time. In
terms of generators this behaviour is described by
\begin{eqnarray}
(LF)(\gamma)&=&\sum_{\{x_1,\ldots,x_n\}\subset\gamma}\int_{\R^d}dy_1\dots\int_{\R^d}dy_n\,
c(\{x_1,\ldots,x_n\},\{y_1,\ldots,y_n\},\gamma)\nonumber\\
&&\cdot\left(F(\gamma\setminus\{x_1,\ldots,x_n\}\cup\{y_1,\ldots,y_n\})-F(\gamma)\right),\label{Dima11}
\end{eqnarray}
where $c(\{x_1,\ldots,x_n\},\{y_1,\ldots,y_n\},\gamma)\geq 0$
indicates the rate at which a group of $n$ particles located at
$x_1,\ldots,x_n$ ($x_i\not=x_j$, $i\not=j$) in a configuration
$\gamma$ hops to the sites $y_1,\ldots,y_n$ ($y_i\not=y_j$,
$i\not=j$). As before, we consider the case
\[
c(\{x_1,...,x_n\},\{y_1,...,y_n\},\gamma)=(KC_{\{x_i\},\{y_i\}})(\gamma\setminus\{x_1,...,x_n\})\geq
0,
\]
where $C_{\{x_i\},\{y_i\}}:=C_{\{x_1,...,x_n\},\{y_1,...,y_n\}}$.
Similar calculations lead then to the expressions
\begin{eqnarray}
&&(\hat LG)(\eta) =\\
&&\i_{\bigsqcup_{k=n}^\infty\Gamma^{(k)}}(\eta)
\sum_{\{x_1,...,x_n\}\subset\eta}
\int_{\R^d}dy_1...\int_{\R^d}dy_n\sum_{\xi\subset\{y_1,...,y_n\}}
\left(C_{\{x_i\},\{y_i\}} \star G(\cdot\cup\xi)\right)
(\eta\setminus\{x_1,...,x_n\})\nonumber\\
&&-\i_{\bigsqcup_{k=n}^\infty\Gamma^{(k)}}(\eta)
\sum_{\{x_1,...,x_n\}\subset\eta}
\int_{\R^d}dy_1...\int_{\R^d}dy_n\sum_{\xi\subset\{x_1,...,x_n\}}
\left(C_{\{x_i\},\{y_i\}}\star
G(\cdot\cup\xi)\right)(\eta\setminus\{x_1,...,x_n\}),\nonumber
\end{eqnarray}
and
\begin{eqnarray}
(\hat L^*k)(\eta)&=&
\int_{\Gamma_0}d\lambda(\zeta)\int_{\Gamma^{(n)}}dm^{(n)}(\xi)
\sum_{\eta_1\subset\eta}k(\zeta\cup (\eta\setminus\eta_1)\cup\xi)\\
&&\cdot\int_{\Gamma_0}d\lambda(\tau)
\i_{\Gamma^{(n)}}(\eta_1\cup\tau)\sum_{\eta_2\subset\eta\setminus\eta_1}
C_{\xi,\eta_1\cup\tau}(\zeta\cup\eta_2)\nonumber\\
&&-\int_{\Gamma_0}d\lambda(\zeta)\int_{\Gamma_0}d\lambda(\xi)\,
k(\zeta\cup\eta\cup\xi)\sum_{\eta_1\subset\eta}\i_{\Gamma^{(n)}}(\eta_1\cup\xi)\nonumber\\
&&\cdot\int_{\Gamma^{(n)}}dm^{(n)}(\tau)\sum_{\eta_2\subset\eta\setminus\eta_1}
C_{\eta_1\cup\xi,\tau}(\zeta\cup\eta_2).\nonumber
\end{eqnarray}
Moreover
\begin{eqnarray}
(\tilde LB)(\theta)&=&
\frac{1}{n!}\int_{\Gamma_0}d\lambda(\eta)\,e_\lambda(\theta+1,\eta)
\int_{\Gamma^{(n)}}dm^{(n)}(\xi)\,
(D^{\vert\eta\vert+n}B)(\theta,\eta\cup\xi)\nonumber\\
&&\cdot\int_{\Gamma^{(n)}}dm^{(n)}(\zeta)\,C_{\xi,\zeta}(\eta)
\big(e_\lambda(\theta+1,\zeta)-e_\lambda(\theta+1,\xi)\big).
\end{eqnarray}

\begin{remark}
If the rate $c$ does not depend on the configuration,
\[
c(\{x_1,\ldots,x_n\},\{y_1,\dots,y_n\},\gamma)=c(\{x_1,\ldots,x_n\},\{y_1,\dots,y_n\}),
\]
one can show that each Poisson measure $\pi_z$, $z>0$, is invariant.
If, in addition, the rate $c(\{x_1,\ldots,x_n\},\{y_1,\dots,y_n\})$
is symmetric in $x_1,\ldots,x_n,y_1,\dots,y_n$, then these Poisson
measures are symmetrizing.
\end{remark}

In particular, the conditions of the previous Remark hold for $n=2$
and
\[
C_{\{x_1,x_2\},\{y_1,y_2\}}=
p(x_1-y_1)p(x_1-y_2)p(x_2-y_1)p(x_2-y_2)e_\lambda(0),
\]
where $p:\R^d\to\R^+_0$ is either an even or an odd function. In this case, 
denoting by $c(x_1,x_2,y_1,y_2)=p(x_1-y_1)p(x_1-y_2)p(x_2-y_1)p(x_2-y_2)$, one 
obtains the following explicit formulas
\begin{eqnarray}
&&\left( \hat{L}G\right) \left( \eta \right)  \\
&=&\i_{|\eta|\geq2}\sum_{\left\{ x,y\right\} \subset \eta }\int_{\mathbb{R}^{d}}dx^{\prime }\int_{%
\mathbb{R}^{d}}dy^{\prime }c\left( x,y,x^{\prime },y^{\prime
}\right) \left[ G\left( \eta \cup \left\{ x^{\prime },y^{\prime
}\right\} \setminus \left\{
x,y\right\} \right) -G\left( \eta \right) \right]  \nonumber\\
&&+2\i_{|\eta|\geq2}\sum_{\left\{ x,y\right\} \subset \eta }\int_{\mathbb{R}^{d}}dx^{\prime }\,G\left( \eta \cup x^{\prime }\setminus \left\{
x,y\right\} \right)\int_{\mathbb{R}^{d}}dy^{\prime } 
c\left( x,y,x^{\prime },y^{\prime
}\right)\nonumber\\
&&-\i_{|\eta|\geq2}\sum_{\left\{ x,y\right\}\subset \eta } \left(G\left( \eta \setminus x\right)+G\left( \eta \setminus y\right)\right)\int_{\mathbb{R}^{d}}dx^{\prime }\int_{\mathbb{R}^{d}}dy^{\prime } c\left( x,y,x^{\prime },y^{\prime
}\right),\nonumber
\end{eqnarray}
and
\begin{eqnarray}
&&(\hat{L}^{\ast }k)(\eta )\\ &=& \i_{|\eta|\geq2}\sum_{\left\{ x,y\right\} \subset \eta }\int_{\mathbb{R}^{d}}dx^{\prime }\int_{%
\mathbb{R}^{d}}dy^{\prime }c\left( x,y,x^{\prime },y^{\prime
}\right) \left[ k\left( \eta \cup \left\{ x^{\prime },y^{\prime
}\right\} \setminus \left\{ x,y\right\} \right) -k\left( \eta \right) \right] \nonumber\\
&+&\sum_{x\in \eta }\int_{\mathbb{R}^{d}}dx^{\prime }\int_{\mathbb{R}%
^{d}}dy^{\prime }\int_{\mathbb{R}^{d}}dy\,c\left( x,y,x^{\prime
},y^{\prime }\right)\left[ k(\eta \cup \left\{ x^{\prime },y^{\prime
}\right\} \setminus x )-k(\eta \cup y) \right].\nonumber
\end{eqnarray}
Additionally,
\begin{eqnarray}
(\tilde{L}B)(\theta ) &=&\frac{1}{2}\,\int_{\mathbb{R}^{d}}dx
\int_{\mathbb{R}^{d}}dy\,
\frac{\delta ^2B(\theta)}{\delta\theta(x)\delta\theta(y)}\int_{\mathbb{R}^{d}}dx^{\prime}\int_{\mathbb{R}^{d}}dy^{\prime}\,c(x,y,x^{\prime },y^{\prime })\\
&&\cdot\, \left[ (\theta (x^{\prime})+1)(\theta (y^{\prime })+1)-(\theta (x)+1)(\theta(y)+1)\right].\nonumber
\end{eqnarray}

\subsection*{Acknowledgments}

This work was supported by DFG through SFB-701 (Bielefeld University) and by
FCT POCI, PDCT, and PTDC, FEDER.

\addcontentsline{toc}{section}{References}
\newcommand{\etalchar}[1]{$^{#1}$}

\end{document}